\documentclass[fleqn,usenatbib]{mnras}

\usepackage{newtxtext,newtxmath}

\usepackage[T1]{fontenc}

\DeclareRobustCommand{\VAN}[3]{#2}
\let\VANthebibliography\thebibliography
\def\thebibliography{\DeclareRobustCommand{\VAN}[3]{##3}\VANthebibliography}

\DeclareSymbolFont{matha}{OML}{txmi}{m}{it}
\DeclareMathSymbol{\varv}{0}{bmisymbols}{"1D}

\defcitealias{whittingham2021}{W21}

\usepackage{graphicx}
\usepackage{upgreek}
\usepackage{lipsum}
\usepackage{stfloats}
\usepackage{amsmath}	
\newcommand{\bs}[1]{\boldsymbol{#1}}
\usepackage[normalem]{ulem} 
\usepackage{hyperref}	
\hypersetup{colorlinks=true,linkcolor=blue,citecolor=blue,filecolor=blue,urlcolor=blue,}

\title[The impact of magnetic fields on galaxy mergers -- II]{The impact of magnetic fields on cosmological galaxy mergers -- II. Modified angular momentum transport and feedback}

\author[Whittingham et al.]{
Joseph Whittingham$^{1,2}\thanks{E-mail: jwhittingham@aip.de (AIP)}$, 
Martin Sparre$^{2,1}$, 
Christoph Pfrommer$^{1}$, 
R{\"u}diger Pakmor$^{3}$
\\
$^{1}$Leibniz-Institute for Astrophysics Potsdam (AIP), An der Sternwarte 16, 14482 Potsdam, Germany\\
$^2$Institut f\"ur Physik und Astronomie, Universit\"at Potsdam, Karl-Liebknecht-Str.\,24/25, 14476 Golm, Germany\\
$^{3}$Max Planck Institute for Astrophysics, Karl-Schwarzschild-Str. 1, 85741 Garching, Germany}

\date{Accepted XXX. Received YYY; in original form ZZZ}

\pubyear{2023}

\begin{document}
\label{firstpage}
\pagerange{\pageref{firstpage}--\pageref{lastpage}}
\maketitle

\begin{abstract}
The role of magnetic fields in galaxy evolution is still an unsolved question in astrophysics. We have previously shown that magnetic fields play a crucial role in major mergers between disc galaxies; in hydrodynamic simulations of such mergers, the Auriga model produces compact remnants with a distinctive bar and ring morphology. In contrast, in magnetohydrodynamic (MHD) simulations, remnants form radially-extended discs with prominent spiral arm structure. In this paper, we analyse a series of cosmological ``zoom-in'' simulations of major mergers and identify exactly \textit{how} magnetic fields are able to alter the outcome of the merger. We find that magnetic fields modify the transport of angular momentum, systematically hastening the merger progress. The impact of this altered transport depends on the orientation of the field, with a predominantly non-azimuthal (azimuthal) orientation increasing the central baryonic concentration (providing support against collapse). Both effects act to suppress an otherwise existent bar-instability, which in turn leads to a fundamentally different morphology and manifestation of feedback. We note, in particular, that stellar feedback is substantially less influential in MHD simulations, which allows for the later accretion of higher angular momentum gas and the subsequent rapid radial growth of the remnant disc. A corollary of the increased baryonic concentration in MHD simulations is that black holes are able to grow twice as large, although this turns out to have little impact on the remnant's development. Our results show that galaxy evolution cannot be modelled correctly without including magnetic fields.
\end{abstract}


\section{Introduction} 
\label{sec:introduction}

Magnetic fields permeate the Universe at every scale yet observed. The galactic scale is, of course, no exception to this. This has been confirmed both for our own galaxy and external galaxies through a wide range of techniques, including Zeeman splitting \citep{heiles2009, li2011, mcbride2015}, stellar light polarisation \citep{heiles1996, pavel2014, berdyugin2014}, dust polarisation \citep{hildebrand1988, lopez-rodrigues2020}, Faraday rotation \citep{manchester1972, han2018}, and synchrotron radiation \citep{dumke1995, krause2009, bennett2013, planck2016}\footnote{See reviews by \citet{beck2015} and \citet{han2017} and references therein for a more comprehensive list of examples.}. The latter is particularly powerfully demonstrated through the far-infrared (FIR) -- radio correlation, which implies volume-filling magnetic fields for a vast range of galaxy sizes, masses, and morphologies \citep{lacki2010, werhahn2021, pfrommer2022}.

Disc galaxies in the local Universe are of particular interest, as observations imply that field strengths in these are on the order of ${\sim}10\,\upmu$G \citep{beck2011}. This places the energy density of the magnetic field in approximate equipartition with the turbulent, thermal, and cosmic ray energy densities in the interstellar medium (ISM) \citep{boulares1990, beck1996, beck2015}, making it a dynamically-important component at the present time. The \textit{long-term} impact of magnetic fields on galactic evolution, however, is still an open question.

To answer this question from a theoretical standpoint, we require the use of cosmological simulations, in which a full range of important environmental factors, such as accretion histories, circumgalactic media (CGM), and mergers, can be accounted for and treated self-consistently \citep{vogelsberger2020}. Many of the latest generation of cosmological simulations now include an implementation of magnetohydrodynamics (MHD), including zoom-in simulations of galaxies such as Auriga \citep{grand2017}, FIRE-2 \citep{su2017}, and those performed by \citet{rieder2017}, as well as larger box simulations such as CHRONOS++ \citep{gheller2016, vazza2017}, Illustris TNG \citep{pillepich2018}, and HESTIA \citep{libeskind2020}. However, the use of different numerical codes, seed fields, and divergence cleaning methods, amongst other factors, has led to inconclusive results. For example, in some simulations of more isolated galaxies, the magnetic field is typically subdominant for the entire runtime \citep{hopkins2020}, whilst in others the magnetic field does reach equipartition, but either only in specific density ranges \citep{ponnada2022} or only at late times \citep{pakmor2017}, thereby limiting its impact. On the other hand, in simulations that start with a sufficiently strong primordial field, magnetic fields are able to suppress star formation rates \citep{marinacci2016} and reduce disc sizes \citep{martin-alvarez2020, katz2021}. The seed strengths used in these simulations are, however, beyond the currently accepted upper limits achievable by standard battery processes \citep{gnedin2000, attia2021}.

In \citet[hereafter \citetalias{whittingham2021}]{whittingham2021}, we pointed out that mergers can raise field strengths to dynamically-important values, even in simulations that start with significantly weaker seed fields. To show this, we ran four pairs of high-resolution cosmological zoom-in simulations of major mergers between disc galaxies, using the Auriga galaxy formation model \citep{grand2017}. We showed that, under this scenario, MHD simulations produce remnants with systematically different sizes and morphologies compared to their hydrodynamic analogues. Specifically, for the merger scenarios we simulated, remnants in the MHD simulations are larger with flocculent gas discs and spiral arms, whilst those in the hydrodynamic simulations are more compact and exhibit conspicuous bar and ring elements.

In the same paper, we also revisited four pairs of high-resolution simulations, originally run by \citet{pakmor2017}. These employ the Auriga model as well, but apply it to galaxies with considerably more quiescent merger histories. Here too, however, similar, albeit more subtle, morphological differences are evident between the MHD and hydrodynamic variants. We interpreted the observation that the differences are stronger in our own simulations as evidence that this is an MHD effect excited by mergers. The observation of similar morphological differences in simulations of more isolated galaxies should not be surprising, though, as few if any galaxies will be untouched by mergers during their history. Indeed, mergers constitute a fundamental part of hierarchical structure formation – a cornerstone of $\Lambda$CDM (a cold dark matter Universe with a cosmological constant).

Significantly, for each of the eight sets of high-resolution simulations mentioned, only the MHD simulations produce galaxies consistent with observations. By analysing kinetic and magnetic energy power spectra for simulations with varying resolution, we demonstrated in \citetalias{whittingham2021} that sufficiently small-scale turbulence must be resolved in order to amplify the  magnetic fields in the necessary time frame and thereby realise this effect. We did not, however, explain \textit{how} the magnetic fields were affecting the re-growth of the disc. We answer this question in this paper.

The paper is organised as follows: in Sec.~\ref{sec:methodology}, we summarise the merger scenarios and our numerical methods. In Sec.~\ref{sec:analysis}, we identify how the mergers evolve differently under hydrodynamic and MHD physics models (Sec.~\ref{subsec:how_evolution_differs}) and propose a mechanism by which magnetic fields are able to cause this effect (Sec.~\ref{subsec:model}). We then provide evidence for this model, with particular emphasis on how magnetic fields affect angular momentum transport and subsequent orbital resonances (Sec.~\ref{subsec:angmom} and Sec.~\ref{subsec:resonances}) and how they alter the manifestation of stellar and black hole feedback (Sec.~\ref{subsec:feedback} and Sec.~\ref{subsec:AGN}). In Sec.~\ref{sec:discussion}, we discuss the applicability of our results to different merger scenarios, numerical codes, and galaxy formation models. Finally, in Sec.~\ref{sec:conclusions}, we summarise our conclusions.

\section{Methodology}
\label{sec:methodology}

In this work, we analyse the four pairs of high-resolution cosmological zoom-in simulations first presented in \citetalias{whittingham2021}. These in turn, are augmentations of the original hydrodynamic simulations presented in \citet{sparre2016, sparre2017} with, in particular, the new additions of Monte-Carlo tracer particles and magnetic fields. The suite consists of four different merger scenarios, with each scenario ran twice from the same initial conditions: once with magnetic fields and once without. In each case, the same underlying numerical implementation is used, such that a hydrodynamic simulation is equivalent to an MHD simulation with the seed field set to zero. We summarise our set-up here, but direct the reader to section 2 of \citetalias{whittingham2021} and references therein for a more comprehensive description.

\subsection{Merger scenarios}
\label{subsec:merger_scenarios}

Each simulation pair focuses on a spiral galaxy that undergoes a gas-rich major merger with another spiral galaxy between redshift $z = 0.9 - 0.5$. The merger mass ratios range between approximately 1.1 and 2 (see table 2 of \citetalias{whittingham2021} for exact details). The mergers may also be roughly separated into in-spiralling (1330, 1526) and head-on (1349, 1605), but cover a variety of impact parameters, speeds, and orbits. Post-merger, the galaxies are allowed to rebuild in relative isolation, experiencing no further events in their merger tree. We note, however, that as these are cosmological simulations, the remnants are not wholly isolated from subsequent minor tidal interactions (see section 2.4 of \citetalias{whittingham2021} for details). By $z = 0$, each remnant is able to rebuild a disc and has a final stellar mass in the range of $6-11\times 10^{10}\; \mathrm{M}_\odot$.

As discussed in section 2 of \citetalias{whittingham2021}, these mergers were specifically chosen with the intention of observing magnetic fields at their most influential; gas-rich progenitors implied strong initial magnetic fields, whilst major mergers were expected to best facilitate field amplification through compression and turbulence. Finally, the disc-rebuilding phase would elongate the time over which the magnetic fields could act. We note, however, that morphological considerations were not part of the original selection criteria \citep{sparre2016, sparre2017}.

We keep the labels for each simulation introduced in \citetalias{whittingham2021}, where a suffix of `H' or `M' represents the inclusion of hydrodynamic or MHD physics, respectively, and the first four digits represent the friends-of-friends \citep{davis1985} group number in Illustris \citep{vogelsberger2014, vogelsberger2014b, genel2014} from which the merger scenario was originally chosen. In order to be consistent with earlier analysis by \citet{sparre2016, sparre2017}, we always present data for the primary progenitor, defined here as the one with the largest stellar mass at $z = 0.93$ \citepalias[see section 2.3 of][]{whittingham2021}. We note, however, that this is a somewhat arbitrary choice, as both of the main progenitors have very similar properties pre-merger, including similar magnetic field strengths out to similar radii.

\subsection{Initial conditions and parameters}

Zoom-in initial conditions were created for each merger scenario using a modified version of the \textsc{N-GenIC} code \citep{springel2015}. In these, a volume of high resolution particles is focused on the target galaxy and its immediate surroundings, with a dark matter mass resolution equal to $1.64 \times 10^5 \; \mathrm{M}_\odot$. This is approximately 38.5 times finer than the original Illustris simulation. A buffer region envelops these particles, with yet coarser resolution particles filling the remaining volume. This volume has a side length of 75 co-moving Mpc~$h^{-1}$.

The softening length used is a co-moving length before $z = 1$, at which point it is frozen at a physical value of 0.22 kpc. For gas cells, this value is also scaled by the cell radius, with the restriction that the minimum softening length is bounded by 30~co-moving~pc~$h^{-1}$ below and 1.1 kpc above. This choice helps to prevent unrealistic two-body interactions at early times, whilst allowing small-scale structure to continue to form at late-times
\citep[see, e.g.,][]{power2003}.

The cosmological parameters were taken from WMAP-9 \citep{hinshaw2013}, with Hubble's constant $H_0 = 100 \,h~\rmn{km~s}^{-1}~\rmn{Mpc}^{-1} = 70.4$ km s$^{-1}$ Mpc$^{-1}$ and the density parameters for matter, baryons, and a cosmological constant, respectively, being $\Omega_\text{m} = 0.2726$, $\Omega_\text{b} = 0.0456$, and $\Omega_\Lambda = 0.7274$.

\subsection{\textsc{Arepo} and Monte-Carlo tracers}
\label{subsec:arepo_mc_tracers}

The simulations were ran from $z = 127$ to $z = 0$ using the moving-mesh code \textsc{Arepo}, which employs a second-order finite-volume Godunov scheme \citep{springel2010, pakmor2016, weinberger2020}. Gas cells in the high-resolution region are refined and derefined so that they stay within a factor of two of the target mass, $2.74 \times 10^4 \; \mathrm{M}_\odot$. Meanwhile, mesh-generating points may be moved arbitrarily. Together, these features allow the code to behave in a quasi-Lagrangian manner, reducing the level of numerical diffusion, whilst simultaneously inheriting the robust nature of grid-based Eulerian codes. The resultant increased accuracy of this method over standard smoothed-particle hydrodynamic (SPH) methods has been well-documented \cite[see, e.g.,][]{vogelsberger2012, sijacki2012, dusan2012}. Of particular importance to this work, is the ability to replicate the Kolmogorov turbulent cascade \citep{kolmogorov1941} for subsonic turbulence, which is not achievable by standard SPH models \citep{bauer2012}. We showed in \citetalias{whittingham2021} that such a cascade was almost certainly crucial to achieving sufficient magnetic field amplification during the merger.

As \textsc{Arepo} is only a quasi-Lagrangian code, to follow the accurate flow of mass in the simulations, we employ the use of Monte-Carlo tracers \citep{genel2013}. We place five tracers per gas cell at the start of each simulation. These are then exchanged with neighbouring gas cells at a rate proportional to the mass flux across their boundaries. Tracers may also be accreted by black holes and be exchanged with star particles during star formation and stellar mass loss processes. Monte-Carlo tracers have been shown to follow the mass flux more accurately than the traditional method of Lagrangian tracers \citep{genel2013}.

\subsection{Auriga}

The galaxy formation physics in the simulations are evaluated using the Auriga model \citep{grand2017}. This model was originally built to produce Milky Way (MW)-like galaxies in zoom-in simulations, and has been able to produce appropriate stellar masses, sizes, rotation curves, star formation rates, and metallicities \citep{grand2017}, the correct structural parameters of bars \citep{calero2019}, and the existence of chemically distinct thick and thin discs \citep{grand2018}. The models for star formation, stellar feedback, and black hole feedback in Auriga are all physically well-motivated and parameters require only limited recalibration between resolution levels\footnote{Whilst parameters must not be significantly retuned, certain phenomena are nevertheless resolution-dependent; for example, the manifestation of starbursts \citep{sparre2016} and the extent of magnetic field amplification post-merger \citepalias{whittingham2021}.}. This is a non-trivial result \citep{scannapieco2012}. We summarise the model below, but encourage the reader to refer to section 2.4 of \citet{grand2017} and references therein for a more complete overview.

\subsubsection{ISM and feedback}
\label{sec:ISM}

The ISM is described by the model of \citet{springel2003}, which assumes that hot and cold phases are in pressure equilibrium and, at the onset of thermal instability, the gas follows a stiff equation of state. In our simulations, this onset (and thus star formation) begins at a threshold gas density of $n_\text{SF}= 0.13\; \text{cm}^{-3}$. The model must not be recalibrated when magnetic fields are introduced \citepalias{whittingham2021}. To replicate Type II supernovae, wind particles are also created out of star-forming gas cells in numbers that reflect the fraction of stars formed in the mass range $8-100\;\text{M}_\odot$. These particles are launched in an isotropically random direction, with a velocity proportional to the local one-dimensional dark matter velocity dispersion \citep{okamoto2010}. Wind particles then interact only gravitationally until they reach a gas cell with $n < 0.05$ $n_\text{SF}$ or exceed a maximum travel time of approximately 25~Myr. At this point they deposit their energy in equal thermal and kinetic parts, which produces smooth, regular winds directed away from the galaxy.

Supermassive black holes are seeded with a mass of $1.4 \times 10^5$~M$_\odot$ once the mass of the corresponding friends-of-friends halo reaches $7.1 \times 10^{10}$~M$_\odot$. Seeding takes place at the position of the most bound gas cell, with dynamics set according to the \citet{springel2005a} model. Black hole accretion takes place predominantly through an Eddington-limited Bondi-Hoyle-Lyttleton model \citep{bondi1944, bondi1952}, with an additional term modelling accretion in the radio mode based on \citet{nulsen2000}. Feedback is implemented through a radio and quasar mode. For the radio mode, bubbles of gas are gently heated at random locations within the halo with a probability following an inverse square profile, whilst for the quasar mode, energy is injected isotropically into the 512 gas cells nearest the black hole. In both cases, energy is injected at a rate proportional to the black hole accretion rate.

\subsubsection{MHD implementation}

Magnetic fields are treated in the ideal MHD approximation \citep{pakmor2011, pakmor2013}, with equations solved using an HLLD Riemann solver \citep{miyoshi2005}. Divergence cleaning is handled using a Powell 8-wave scheme \citep{powell1999}. This scheme has been found to be more robust than the competing Dedner \citep{dedner2002} scheme when applied to cosmological simulations \citep{pakmor2013}. Our MHD implementation can replicate a variety of phenomena, including: the linear phase of growth of the magneto-rotational instability \citep{balbus1991, pakmor2013, zier2022}, the correct propagation of Alfv\'en and magnetosonic waves in co-moving coordinates \citep{berlok2022}, the development of a small-scale dynamo in MW-like galaxies \citetext{\citealt{pakmor2014}, \citealt{pakmor2017}, \citetalias{whittingham2021}, \citealt{pfrommer2022}},
 similar field strengths and radial profiles to those observed in MW-like galaxies \citep{pakmor2017}, and Faraday rotation measure strengths that are broadly consistent to those observed for MW-like galaxies, both for the disc \citep{pakmor2018} and when compared with the current upper limits available for the CGM \citep{pakmor2020}.

We seed magnetic fields in our initial conditions with a strength of $10^{-14}$ co-moving Gauss. This choice is essentially arbitrary, as the initial configuration and field strength is quickly erased for a broad range of values in collapsing haloes \citep{pakmor2014}. This seed strength is also dynamically irrelevant outside of collapsed haloes \citep{marinacci2016}. Magnetic energy is assumed to be locked up in wind- and star-forming events, but is otherwise not explicitly included in our subgrid models.

\section{Analysis}
\label{sec:analysis}

\subsection{How the evolution of the merger remnant differs between physics models}
\label{subsec:how_evolution_differs}

\begin{figure*}
    \includegraphics[width=\textwidth]{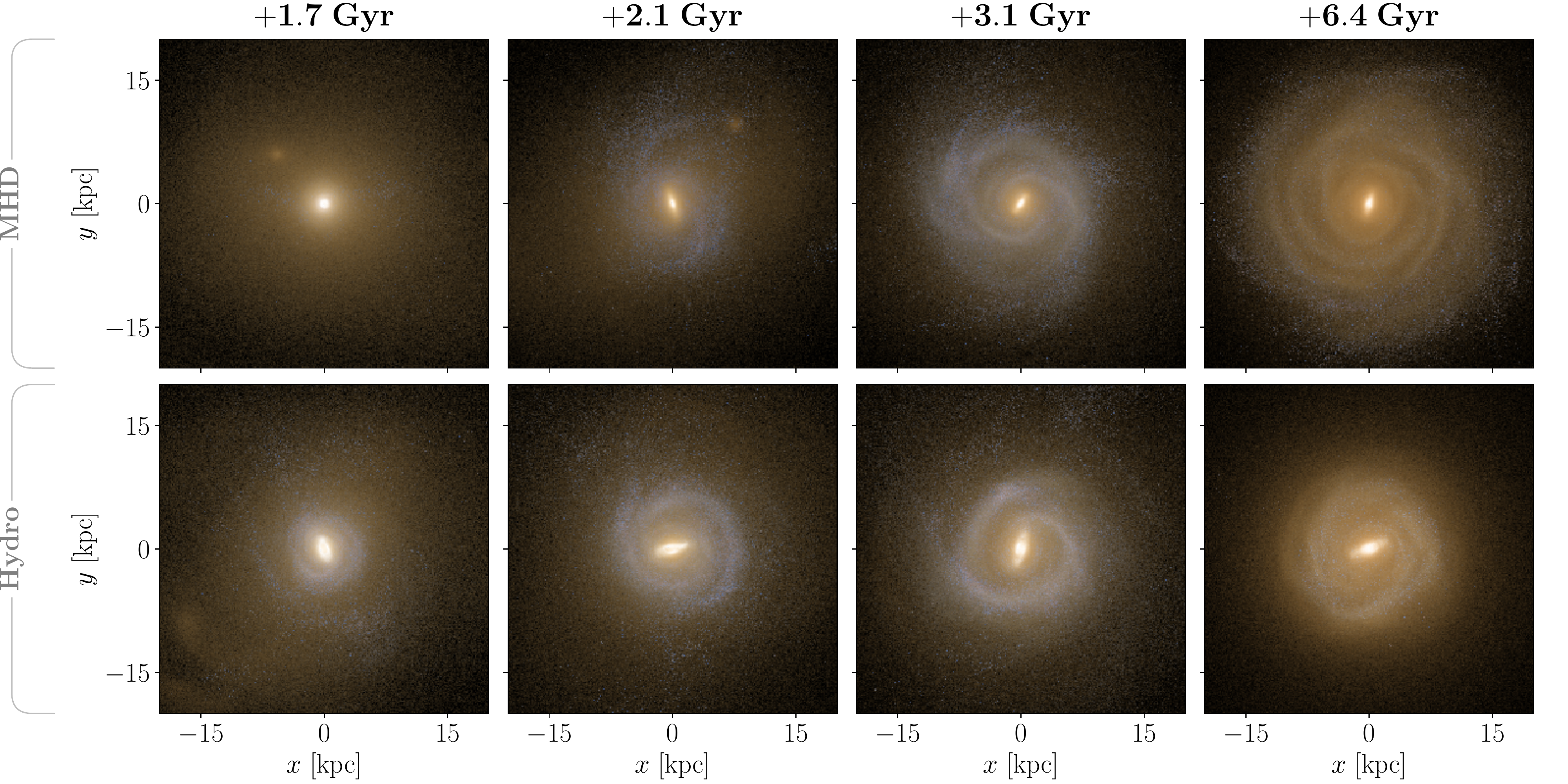}
    \caption{\textit{Top row:} Mock \textit{gri} composite images showing the evolution of the remnant in the 1349-3M simulation post-merger. \textit{Bottom row}: As above, but for the 1349-3H hydrodynamic simulation. Labels above each column indicate time elapsed since first closest approach. The formation of a strong bar in the hydrodynamic simulation is associated with the development of a stellar ring. The absence of a bar in the MHD simulation is associated with the formation of more varied small-scale structure. An animated version of this figure can be found \href{https://youtu.be/C9IAcu5G8Es}{here}.}
    \label{fig:mock_visual}
\end{figure*}

We start our analysis by isolating exactly how the evolution of the merger remnants differs for hydrodynamic and MHD simulations. We will use the 1349-3 simulations here as a case study, being broadly representative of the wider simulation suite. We will also focus on the stellar light morphology, which better highlights the evolution of the distinctive bar, ring, and spiral arm components. To this end, in Fig.~\ref{fig:mock_visual}, we present a series of face-on mock \textit{gri} images\footnote{The differences between the edge-on images are more subtle, and so we defer analysis of these to Appendix~\ref{appendix:edge-on-gri-images}.}. These images were created in the same manner as described in \citetalias{whittingham2021}, using the photometric properties of all star particles within $\pm30$ kpc of the midplane. For each snapshot, the time elapsed since the beginning of the merger (defined here as the time of first closest approach) is given above each column. We have chosen times such that each snapshot shows a significant step in the evolution of the remnant morphology, with the last column equivalent to $z=0$.

As stated in Sec.~\ref{subsec:merger_scenarios}, each of our simulated remnants is able to reform a disc post-merger. However, whilst the remnant in the hydrodynamic simulation starts to rebuild a disc almost immediately, this process is initially delayed in the MHD simulation. This leads to a substantial difference in the size of the respective discs, as observed in the leftmost column of Fig.~\ref{fig:mock_visual}. Once the disc rebuilding process in the MHD simulation begins in earnest, however, progress is rapid. Indeed, the radial size of the disc in the MHD simulation ultimately outstrips that of its hydrodynamic analogue, as can be seen in the final column of the figure.

As well as the size evolution, the structural evolution of each remnant also differs; even at the earliest snapshot shown, in the hydrodynamic simulation a distinct bar and ring morphology is apparent. This ring is star-forming, as can be determined from its bluish hue, which reflects a young stellar population. The ring reaches a fairly steady form already by the second snapshot, with growth plateauing shortly thereafter. On the other hand, the bar formed in the MHD simulation is substantially weaker, and, instead of a ring, a substantial amount of small-scale structure is formed. This small-scale structure at first takes the form of distinct spiral arms before the stellar distribution eventually becomes more flocculent. In the final snapshot shown, the colours in both sets of \textit{gri} images become more yellow as the bulk of star formation has finished and the luminosity is now dominated by older stars. For the hydrodynamic simulation, this is associated with the ring structure becoming less well-defined. 

The differences observed in Fig.~\ref{fig:mock_visual} prompt three important questions, upon which we will base the analysis in this paper. These are:

\begin{enumerate}
    \item Why is the stellar population initially so much more compact in the MHD simulation?
    \item Why does a bar and ring structure form in the hydrodynamic simulation but not in the MHD simulation?
    \item Why does the remnant in the hydrodynamic simulation reach a maximum size, whilst that in the MHD simulation continues to grow?
\end{enumerate}

\subsection{Model for how magnetic fields affect mergers}
\label{subsec:model}

\begin{figure*}
    \includegraphics[width=\textwidth]{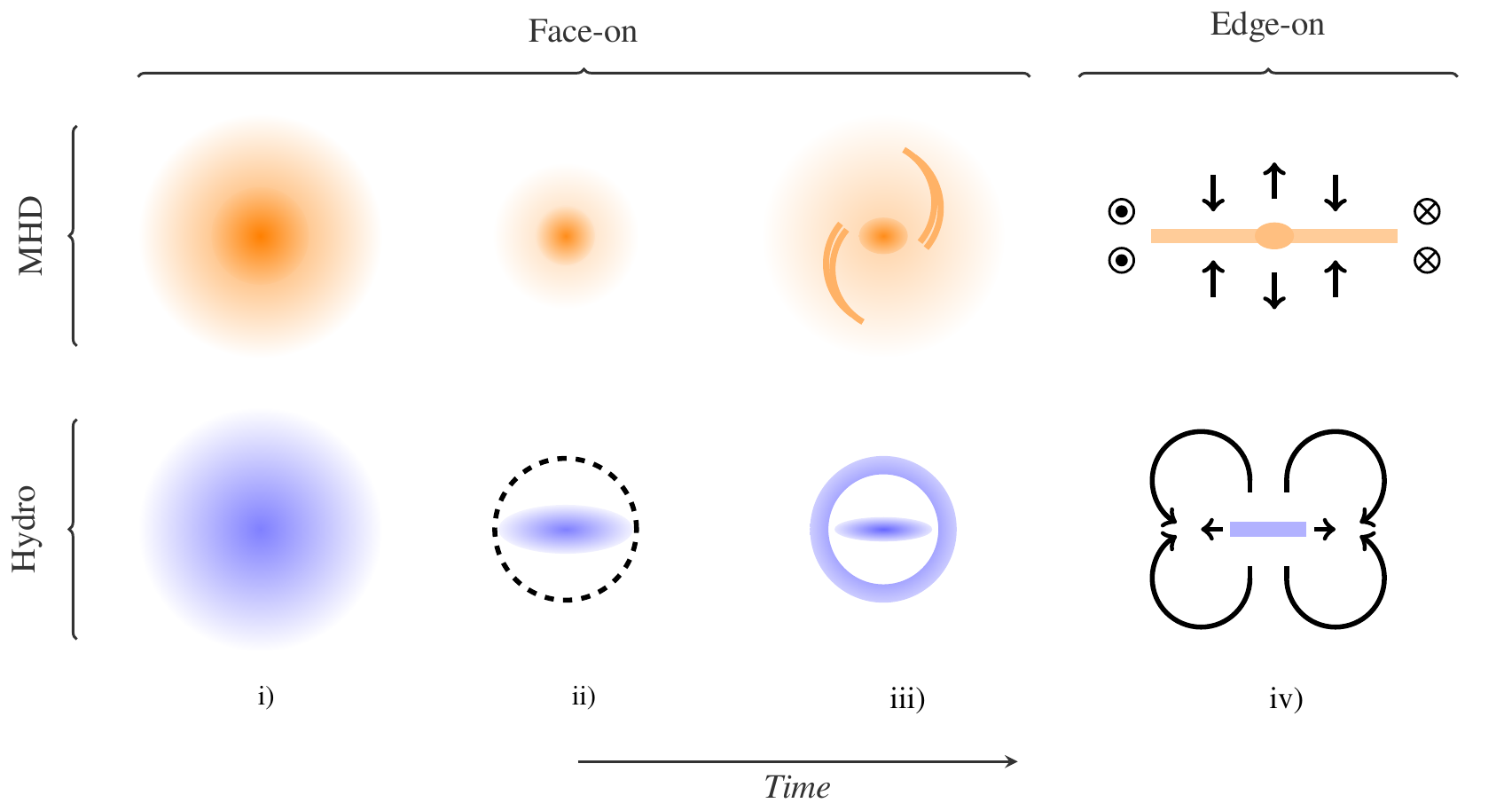}
    \caption{A schematic illustrating the key stages of development in our MHD and hydrodynamic simulations post-merger. Amplified magnetic fields are able to mediate angular momentum, which typically increases the baryonic concentration, thereby suppressing a bar instability. This leads to a fundamentally different stellar distribution and manifestation of feedback. A full description of each stage can be found in Sec.~\ref{subsec:model}.}
    \label{fig:schematic}
\end{figure*}

Cosmological simulations are intrinsically complicated by their very nature. Accordingly, there are several factors that must be taken into account when explaining how magnetic fields are able to affect the outcome of mergers. To simplify our explanation, we first outline a streamlined model of the stages involved before presenting evidence for each of these stages in the remainder of the paper. We present a visual representation of the stages in Fig.~\ref{fig:schematic}, with descriptions given below:

\begin{enumerate}
 \item \textit{Angular momentum transfer:} The merger significantly amplifies the magnetic field, allowing it to have a strong dynamical back-reaction on the gas. This typically takes place within a few 100~Myr of the first closest approach. When the magnetic field is non-azimuthally orientated, the redistribution of angular momentum between gas cells is more effective, leading to a loss in total angular momentum in the disc and a subsequently higher central baryonic concentration.
 \item \textit{Suppression of a bar instability:} In the hydrodynamic simulations, the post-merger starburst causes the remnant to become bar-unstable. This instability is suppressed in the MHD simulations. The exact cause of this suppression depends on the magnetic field configuration; when the field is predominantly non-azimuthally orientated, the suppression originates from the generation of a strong inner Lindblad resonance caused by the increased mass concentration. In the azimuthal case, the magnetic field suppresses the instability by providing support against collapse.
 \item \textit{Resonances:} The large bar in the hydrodynamic simulation reorders the existing distribution of stars and shepherds gas towards the outer Lindblad resonance. This causes an exceptionally high star formation rate in this region. The absence of a strong bar in the MHD simulations allows the gas to remain flocculent and for spiral arm features to develop.
 \item \textit{Winds:} The high star formation rate density in the hydrodynamic simulation launches a strong stellar wind. This acts both radially away from the disc and initiates a large-scale fountain flow. Together, these winds strongly disrupt the angular momentum of accreting gas, helping to keep the disc compact. In the MHD simulations, star formation is more spread out, and stellar winds consequently have a much lower impact. Indeed, at the outskirts of the remnant, gas can be almost co-rotating, allowing it to join the disc practically in-situ. This facilitates rapid radial growth.
\end{enumerate}
The result of these steps is that the remnant in the MHD simulation forms a typical spiral galaxy with an extended radial profile, whilst in the hydrodynamic simulation, the remnant is substantially smaller and displays prominent bar and ring components. 

We present evidence for this model in the following sections. We focus on the \textit{Angular momentum transfer} stage in Sec.~\ref{subsec:angmom}, on the \textit{Suppression} and \textit{Resonance} stages in Sec.~\ref{subsec:resonances}, and on the \textit{Winds} stage in Sec.~\ref{subsec:feedback}.

\subsection{How magnetic fields increase the baryonic concentration through modified angular momentum transport}
\label{subsec:angmom}

\begin{figure*}
    \includegraphics[width=\textwidth]{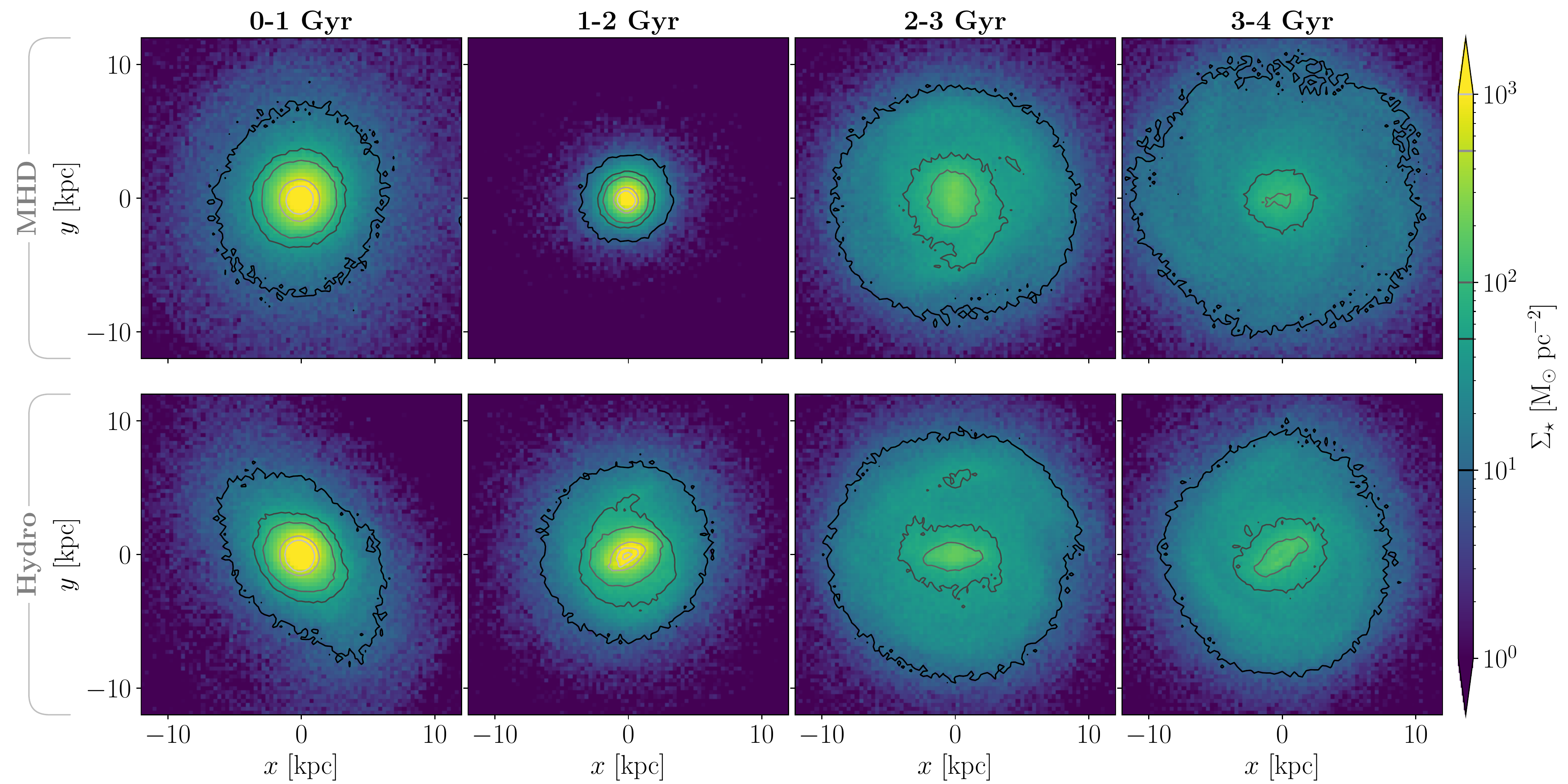}
    \caption{\textit{Top row:} Stellar surface density maps for 1349-3M, where stars have been selected such that they were formed in the previous Gyr. Maps show the distribution at +1, +2, +3, and +4 Gyr post-merger, respectively. Contours are shown at 10, 50, 100, 500, and 1000 $\text{M}_\odot\; \text{pc}^{-2}$. The projection has a vertical extent of $\pm5$ kpc from the midplane. \textit{Bottom row:} As above, but for 1349-3H. Star formation between 1 and 2 Gyr post-merger is more concentrated in the MHD simulation. This stabilises the disc against the formation of a bar. In contrast, in the hydrodynamic analogue, a large bar forms. This leads to a markedly different stellar distribution.}
    \label{fig:bar_star_ages}
\end{figure*}

To illustrate how angular momentum in the disc evolves differently between the two physics models, we start by tracing how and where stars form in the successive Gyrs post-merger. This is directly affected by how the dense gas is distributed, upon which the magnetic fields have an influence. To this end, in Fig.~\ref{fig:bar_star_ages} we show stellar surface density maps for the 1349-3 simulations, where in each panel we have selected only the stars that formed in the previous Gyr. That is to say, the first column is shown at 1~Gyr post-merger (equivalently, 1~Gyr after first closest approach) and includes stars formed between 0 and 1~Gyr post-merger, the second is shown at 2 Gyr post-merger and includes stars formed between 1 and 2~Gyr post-merger, and so forth. By binning the star formation over this time interval, we smooth over the inherent stochasticity of our underlying star formation model. This, of course, leaves up to a Gyr for the stars to move from their birth position, but in practise, we observe that migration during this time is limited.

In the first column, the distributions are approximately isotropic in both cases. This isotropy is especially strong in the case of the MHD simulation. In the hydrodynamic simulation, the distribution becomes slightly skewed as we move away from the centre. This is principally a projection effect; at the time this surface density map is made, the disc is reorientating in space as material with different orbital angular momenta is accreted. Newly-born stars at the outskirts of the disc have not yet reorientated to orbit in the plane perpendicular to the line of sight. The lack of such an effect in the MHD simulation results from angular momentum transfer facilitated by the magnetic field, which acts to keep the disc rotating coherently. We will show evidence for this in the following plot.

By the second column of Fig.~\ref{fig:bar_star_ages}, there are already noticeable differences between the two remnants. Most strikingly, the stellar population in the MHD simulation is now significantly more compact, whilst the distribution in the hydrodynamic simulation remains extended, as we saw previously in the stellar light distribution in Fig.~\ref{fig:mock_visual}. Both remnants have formed roughly the same amount of stars at this point \citepalias{whittingham2021}, implying a stellar concentration that is significantly higher in the MHD case\footnote{We will show this is true from a more quantitative standpoint in Sec.~\ref{subsec:suppression}.}. Indeed, whilst the surface mass density increases towards the centre in the MHD case, the innermost contour in the hydrodynamic analogue actually marks the reduction of the surface density below 1000~M$_\odot\;\text{pc}^{-2}$ again. This reduction is a typical response of gas to a bar potential \citep{kormendy2004}. The existence of this bar can be seen more explicitly through the increased anisotropy of the innermost contours, as well as implicitly through the faint outline of a stellar lane traced out by the 50~M$_\odot\;\text{pc}^{-2}$ contour. As we will see later in Sec.~\ref{subsec:stellar_ring}, the tidal impact of the bar is critical for producing the associated ring-shaped morphology in hydrodynamic remnants.

By +3 Gyr, the majority of the post-merger star formation has finished. This is reflected by the fact that stellar surface density contours in the third and fourth columns of Fig.~\ref{fig:bar_star_ages} only exist up to 100~M$_\odot\;\text{pc}^{-2}$. Nonetheless, some morphological evolution continues to take place in these panels. Firstly, it can be seen that the bar in the hydrodynamic simulation continues to develop and is supported by fresh star formation. With a keen eye, the faint traces of a star-forming ring can also be seen, close to the outermost contour\footnote{Explicit evidence of this feature will be shown in Sec.~\ref{subsec:stellar_ring}.}. In the MHD case, on the other hand, the innermost contours become slightly more anisotropic with time, as the magnetic field dominance weakens, and the beginnings of spiral arms start to appear instead of a ring (cf. the features in Fig.~\ref{fig:mock_visual}). 

In both simulations, the evolutionary step between 1 and 2~Gyr post-merger is key to the final outcome; in the hydrodynamic simulation, a bar starts to form during this time, which goes on to have a strong tidal impact on the rest of the remnant. In contrast, in the MHD simulation, the disc appears to be stabilised against bar formation during this time through its compaction. Compaction to this extent requires a substantial reduction in the average magnitude of the gas angular momentum. This is, in turn, a direct result of the mediation of angular momentum by the magnetic field, as we show in the following figure.

\begin{figure*}
    \includegraphics[width=\textwidth]{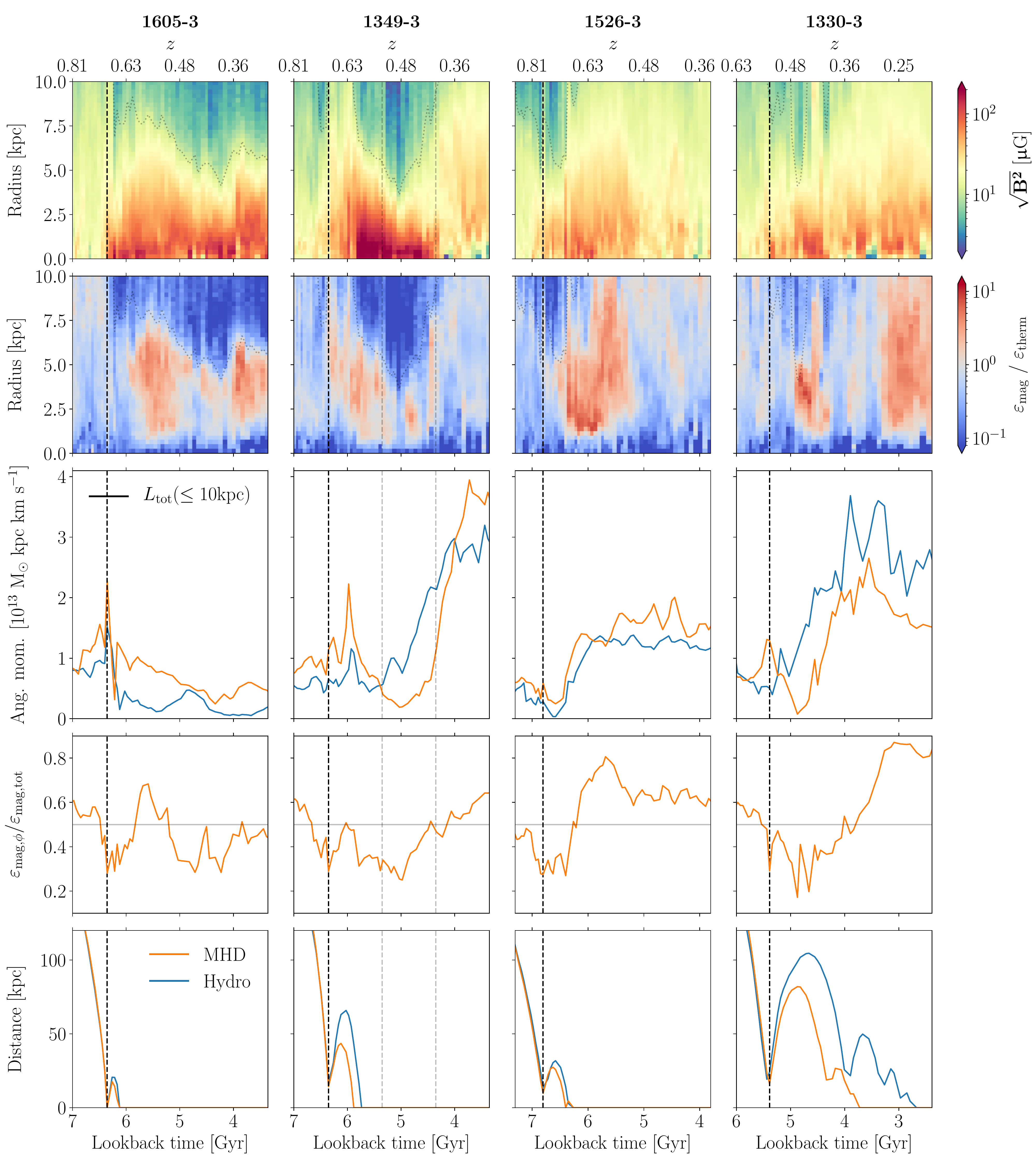}
    \caption{\textit{Top row:} Radially-binned mean magnetic field strength of the main galaxy as a function of time, using annular rings of width 0.25~kpc  and depth $\pm 1$ kpc from the midplane. A dotted line indicates the point at which the gas density drops below $0.02\,\mathrm{M}_\odot\,\mathrm{pc}^{-3}.$ \textit{2nd row:} As above, but showing the magnetic to thermal energy ratio in each ring. \textit{3rd row:} The absolute value of the total angular momentum for all gas cells within a sphere of radius 10~kpc. \textit{4th row:} The fraction of magnetic energy density in the azimuthal component, calculated for a disc bounded by radius 10 kpc and height $\pm 1$ kpc from the midplane. \textit{Bottom row:} The distance between the two merging galaxies as a function of time. The dashed, black, vertical lines mark the time of first closest approach in each simulation. The dashed, grey, vertical lines mark +1 and +2 Gyr post-merger in 1349-3M to aid comparison with Fig~\ref{fig:bar_star_ages}. The merger-induced amplification of the magnetic field allows it to become dynamically important. When the magnetic field is predominantly non-azimuthally orientated, this leads to a more efficient redistribution of gas angular momentum between the accreting gas and that already in the disc. This, in turn, causes an initial reduction in disc size. The same mechanism hastens the progress of the merger in each example.}
    \label{fig:ang_mom}
\end{figure*}

In order to show that the magnetic fields are capable of mediating angular momentum, we first need to show that they are dynamically relevant. With this in mind, in the first row of Fig.~\ref{fig:ang_mom}, we show the mean magnetic field strength in the disc as a function of radius and time. This was created in the same manner as for figure 2 in \citetalias{whittingham2021}, using annular rings of width 0.25~kpc and a vertical extent of $\pm$1 kpc, where this volume is orientated according to the angular momentum vector of the cold gas disc. We focus on a time period for each galaxy that extends from 0.5~Gyr before the start of the merger (marked by the dashed, black, vertical lines) until 3 Gyr afterwards. The colour bar ranges from 2 to 200 $\upmu$G. This scaling covers all but the very innermost radii in 1349-3M during its period of most intense amplification. For this galaxy, the mean field strengths reach a maximum of $310$~$ \upmu$G.

As we noted in \citetalias{whittingham2021}, the first closest approach is associated with a rapid amplification of the magnetic field in all cases. An additional boost also takes place at further passages and at coalescence. This is caused by the additional compression and turbulent injection that takes place at these times. For every galaxy except 1526-3M, there are periods in which the radial extent of the amplified region reduces. This is a signature of the increase in concentration that takes place in these remnants. Such an effect can be observed, for example, for 1605-3M from $5.5-4$~Gyr, starting again at 4~Gyr; for 1349-3M from approximately $6-4$~Gyr; and for 1330-3M from $5-4.5$~Gyr. For aid of comparison between the data presented here and that in Fig.~\ref{fig:bar_star_ages}, we have added dashed, grey, vertical lines to the 1349-3M column to indicate +1 and +2~Gyr post-merger. The reduction in size of the amplified region here is clearly reflected by the reduced size of the stellar distribution in Fig.~\ref{fig:bar_star_ages}. For 1605-3M and 1349-3M, it is also reflected by the general decrease in the radius of the gas disc, as delineated by the dotted line. We use a density threshold of $0.02\,\mathrm{M}_\odot\,\mathrm{pc}^{-3}$ to measure this. All galaxies eventually experience at least a temporary regrowth of this disc as gas accretes. The concentration of the gas can, however, continue to increase during these times if the magnetic field remains strong enough. This is evident, for example, in 1349-3M at a lookback time of around $5-5.5$~Gyr.

In the second row of Fig.~\ref{fig:ang_mom}, we show the mean magnetic to thermal energy density over the same volumes as above. It can be seen that the magnetic energy density within the disc pre-merger is comparable to, if slightly lower than, the thermal energy density. However, within a short time of the merger, the magnetic energy density soon dominates. The balance between the two energy densities then fluctuates due to the back-reaction of the magnetic fields on the gas and the additional injection of turbulence by inflows\footnote{We have also analysed the magnetic to turbulent energy density, using the definition given in eq. 6 in \citet{pakmor2017}, and see similar trends, but with weaker dominance of the magnetic fields. For the magnetic to rotational energy density, magnetic fields are able to reach a similar order of magnitude when dominant in Fig.~\ref{fig:ang_mom}, but are typically a factor of a few weaker.}. The periods in which the magnetic field is dominant in each simulation are also generally reflected by a period of time in which the gas concentration increases. This, in turn, is associated with a decrease in the overall angular momentum in the disc, as we show explicitly in the next row.

In the third row of Fig.~\ref{fig:ang_mom}, we show how the magnitude of the total gas angular momentum within 10 kpc of the remnant, $L=|\bs{L}|$, evolves as a function of time for the MHD (orange) and hydrodynamic (blue) simulations, respectively. We calculate this as $\bs{L}=\Sigma_i (\bs{r}_i \bs{\times} \bs{p}_i)$, where $r_i$ is the radial distance of gas cell, $i$, from the galaxy centre and $p_i$ is its momentum. We evaluate this sum across a sphere to avoid rotating the reference frame, as was done in the upper two rows of the figure. This prevents us contaminating the sum with artificial torques.

It can be seen that in three out of four cases (i.e. all except 1526-3) the evolution of the total angular momentum differs substantially between the MHD simulation and its hydrodynamic analogue. For these, in the MHD simulation, the first closest approach (marked by the dashed, vertical lines) is always associated with a spike in the angular momentum. This is indicative of gas being brought into the 10 kpc radius by the merging galaxy. In the more head-on mergers (1605-3 and 1349-3) the total angular momentum drops shortly afterwards, as gas temporarily leaves the sphere again. This is then followed by a second spike at coalescence as the gas reaccretes. Similar temporary increases in the total angular momentum can usually be seen in the hydrodynamic simulations as well, but these are firstly not always evident and secondly, when they do exist, the spike peaks at systematically lower values. 

This behaviour can be explained by inspecting the distribution of angular momenta components\footnote{We do not show this here due to space constraints.}. For these, we observe that, in the MHD simulations, gas flows reaching the sphere are typically able to remain more coherent. This increases the ability of both matter and angular momentum to penetrate the sphere and reach the galaxy, thereby providing the larger spikes seen in total angular momentum in Fig.~\ref{fig:ang_mom}. The increased coherence of such flows is likely to be a result of them being less easily broken apart due to magnetic draping \citep{dursi2008,berlok2019}, as has been observed, for example, in simulations of jellyfish galaxies passing through the intergalactic medium \citep{sparre2020,mueller2021}. This process would also explain why the gas in 1605-3M exhibits a fairly high degree of angular momentum post-merger, whilst in 1605-3H the angular momentum reduces substantially; gas flows in this merger are strongly misaligned and therefore, in the latter, rapidly become turbulent, whilst they are shielded to a degree from this process in the MHD simulation. We note that this effect can only be realised with sufficiently high resolution.

After an initial increase in the total angular momentum in 1605-3M, 1349-3M, and 1330-3M, this quantity undergoes a sustained decline in these simulations as gas with misaligned angular momentum is accreted and is redistributed amongst the existing material. This decline is a direct measurement of the reduction in disc size of the remnants and clearly corresponds to the signatures already analysed for the upper two rows of the figure. Once again, for 1349-3M, a comparison can be made between Figs.~\ref{fig:bar_star_ages} and \ref{fig:ang_mom} with the aid of the dashed, grey, vertical lines. The reduction in magnitude of the gas angular momenta directly leads to a stellar distribution with lower angular momenta, thereby increasing the stellar concentration relative to its hydrodynamic analogue, as was observed in Fig.~\ref{fig:bar_star_ages}.

For 1349-3M and 1330-3M, sufficient angular momentum is eventually accreted such that the disc starts to grow rapidly again. For 1349-3M, this early disc-regrowth phase also provides further evidence of torquing by magnetic fields, as we show in Appendix~\ref{appendix:edge-on-gri-images}. In 1605-3M, the CGM is too disturbed by outflows (see Sec.~\ref{subsec:feedback} and Appendix~\ref{appendix:stellar_feedback}) to provide any substantial growth, and subsequently the disc continues to mostly decrease in size, save for a brief increase at $\sim$4~Gyr. Similar outflows are likely to stop the growth of the disc in 1605-3H, which experiences a degree of accretion-driven growth post-merger between a lookback time of 5.5 and 5 Gyr before decreasing again\footnote{We note that this decrease also correlates with a period of increased AGN activity (see Sec.~\ref{subsec:AGN}).}.

We have so far neglected 1526-3M, as it does not fit the general pattern; here, even when the magnetic field is dominant, no disc size reduction takes place. This behaviour can be understood, however, by examining the magnetic configuration in this remnant. In the fourth row of Fig.~\ref{fig:ang_mom}, we show the fraction of the magnetic energy density in the azimuthal component, where we have calculated this value using the same volumes analysed in rows 1 and 2. The grey, horizontal line marks the point at which the azimuthal component dominates over the non-azimuthal (i.e. vertical and disc-like radial) components. By comparing rows 2 and 4 of Fig.~\ref{fig:ang_mom}, it can be seen that for 1605-3M, 1349-3M and 1330-3M, the magnetic field experiences sustained periods of dynamical dominance when the field is also non-azimuthally orientated, whilst in 1526-3M, the magnetic field becomes strongly azimuthal just as it also becomes dominant. We believe that this explains why the other three MHD simulations increase their baryonic concentration, whilst 1526-3M does not.

The orientation of the magnetic field is important because of its implications for angular momentum transfer; when the magnetic field is predominantly non-azimuthally-orientated, field lines connect the disc to the CGM. In general, there will be a difference in angular velocity between these two components, which results in a magnetic tension force acting on them. When the gas in the disc is rotating faster, as is typically the case, this tension force decreases the speed of the gas in the disc whilst increasing it in the CGM. Angular momentum is thereby transported out of the disc, shrinking it. This effect will be still stronger if the infalling gas rotates oppositely to that in the disc, as then a drag force applies to both parts. Such a case will generally arise in a turbulent CGM, but will be especially influential in retrograde mergers where the majority of new material is counter-rotating relative to the existing gas disc. This is exactly the case in 1349-3M and 1330-3M, and likely the cause of the large drops seen in their total angular momentum. In 1526-3M, meanwhile, the magnetic field is predominantly azimuthally-orientated. In this case, field lines connect gas cells with similar angular momenta, which limits the impact that angular momentum redistribution can have. This results in a very similar evolution in the total angular momentum for 1526-3M and 1526-3H. However, here too, the magnetic fields have an impact, as, in connecting similar angular momenta gas, the field lines actively isolate the gas from external influences. Consequently, in 1526-3M, the magnetic field does not increase the baryonic concentration, but rather supports the disc against collapse. This encourages more isotropically distributed star formation, which also helps to stabilise the disc against bar formation \citep{sellwood2014}.

As well as affecting the baryonic concentration, the mediation of angular momentum through the magnetic fields also has a larger-scale effect. We explore this in the final row of Fig.~\ref{fig:ang_mom}, where we show the distance between the centres of the two merging galaxies as a function of time. We define the centre of each galaxy as the particle with the lowest potential in the subhalo found by $\textsc{subfind}$ \citep{springel2001}. Coalescence is then defined when $\textsc{subfind}$ can no longer identify two gravitationally ``self-bound'' subhaloes (see section 2.3 of \citetalias{whittingham2021} for further details). It can be seen that the mergers in the MHD simulations coalesce systematically faster than their hydrodynamic analogues. 

The difference in time required for coalescence is greatest in absolute terms when the merger took longest. The 1330-3 simulations are a particularly strong example of this, with the MHD simulation coalescing over a Gyr earlier than its hydrodynamic analogue. The trajectories in this case are practically identical for each pair of galaxies until first closest approach, at which point the merging galaxy in the MHD simulation loses a significant amount of angular momentum. A similar angular momentum transfer also takes place at the next two closest approaches, further quickening the rate of coalescence.

\subsection{The impact of resonances}
\label{subsec:resonances}
\subsubsection{The suppression of a bar in MHD simulations}
\label{subsec:suppression}

\begin{figure*}
    \includegraphics[width=\textwidth]{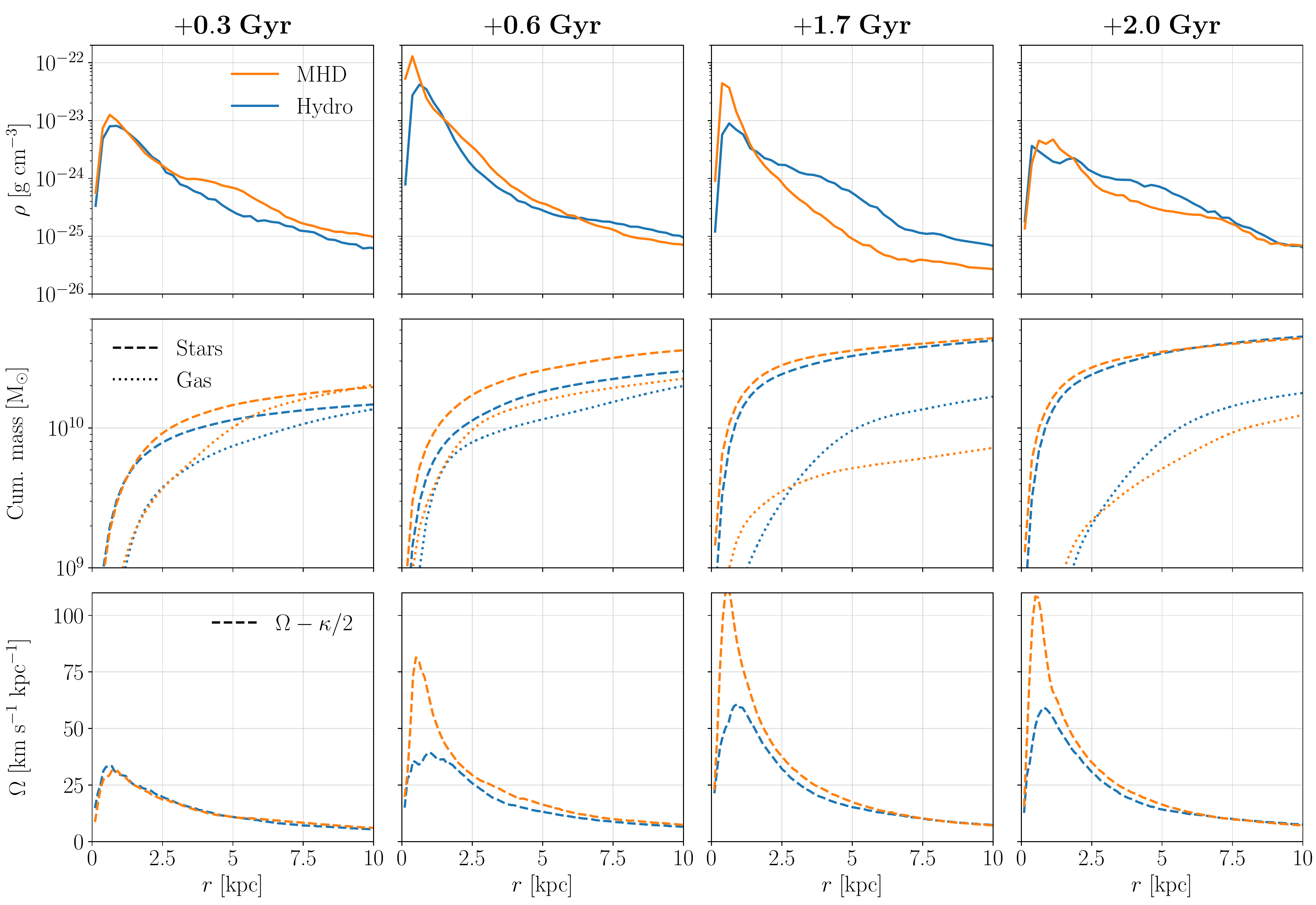}
    \caption{\textit{1st row:} Mean gas density as function of radius, measured in spherical shells of width 0.25 kpc for the 1349-3 simulations. \textit{2nd row:} The cumulative mass in gas and stars, calculated using the same shells as above. \textit{3rd row:} The evolution of the inner Lindblad resonance profile over time. Labels above each column indicate time elapsed since the start of the merger. The increased concentration of gas in the MHD simulation results in a more concentrated distribution of stars. This, in turn, generates a strong inner Lindblad resonance, which acts as a barrier to bar formation. In the hydrodynamic simulation, the peak of the resonant profile is low enough to be overcome and subsequently a strong bar is able to form.}
    \label{fig:gas_concentration}
\end{figure*}

It may perhaps sound contradictory that magnetic fields act to reduce disc sizes immediately post-merger, but lead to larger sizes overall by $z = 0$. However, the compaction stage is actually critical to the remnant's future growth, having a major impact on how resonances form in the disc. To show this, in Fig.~\ref{fig:gas_concentration}, we analyse the relationship between the baryonic concentration in the disc and the subsequent formation of resonances in the 1349-3 simulations. We do this for the first 2 Gyr of the remnant's regrowth, as we identified this as a critical stage in the remnant's development in Sec.~\ref{subsec:angmom}.

In the first row of Fig.~\ref{fig:gas_concentration}, we show radial density profiles of the gas for both MHD and hydrodynamic simulations, calculated using spherical shells of width 0.25~kpc. We previously asserted that the addition of magnetic fields leads to a higher baryonic concentration in the remnant, and we may use Fig.~\ref{fig:gas_concentration} to reevaluate this assertion from a more quantitative standpoint. For the first column, at +0.3 Gyr, we observe that the radial profiles for the inner 2.5 kpc of both physics models are very similar. At distances further out than this, it can be seen that there is a ``bump'' of higher density gas in the MHD case. The timing of this snapshot matches the angular momentum peak seen at coalescence in Fig.~\ref{fig:ang_mom}. This gas likely belongs to the inflows providing the extra angular momentum seen at this time.

In the following panel, the radial profiles begin to look more similar at radii $\gtrsim$2.5~kpc, but a strong peak in the gas density can be seen at the inner central kpc for the MHD simulation. This region is within the range affected by quasar feedback and hence is subject to a certain degree of variability as gas is expelled and then flows back following successive outbursts. The profiles we show here, however, are typical for all following times until approximately +1.7~Gyr post-merger, as shown in the third panel. That is, in the MHD simulation, the gas density outside the 2~kpc region decreases as the disc size reduces, whilst the peak gas density levels are maintained at levels typically several factors higher than in the hydrodynamic analogue.

By +2~Gyr post-merger, as seen in the final column, the peak gas densities are once again similar for both physics models. At this time, the period of most significant amplification has finished, as has the bulk of the starburst. The overall amount of stars formed during this time is approximately the same, as can be seen by inspecting the second row of Fig.~\ref{fig:gas_concentration}, where the cumulative stellar mass profiles at +2 Gyr at distances $\gtrsim$5~kpc approximately match. The distribution of stellar mass, however, is different for each physics model; in the MHD simulation, more mass is found closer to the disc centre. This divergence may appear subtle, but this change in mass concentration has a strong influence on the generation of resonances, which influence the likelihood of bar formation.

In the final row of Fig.~\ref{fig:gas_concentration}, we present profiles for the inner Lindblad resonance. This resonance forms a crucial part of our current understanding of both bar formation and orbital dynamics in barred galaxies \citep[see, e.g.][]{friedl1993, weinberg2007, athanassoula2013, sellwood2014, renaud2015}. For approximately axisymmetric potentials, as inferred from Fig.~\ref{fig:bar_star_ages}, we may calculate the profile of this resonance by employing the epicyclic approximation \citep{binney2008}. Under this, orbits can be considered to be mostly circular with a small radial oscillation about a guiding centre. The frequency of this oscillation, $\kappa$, resonates if it is a multiple of the bar pattern speed, $\Omega_\text{p}$ (the angular frequency at which the bar rotates). We may write this condition as: $m (\Omega_\text{p} - \Omega) = l\kappa$, where $l$ and $m$ are integers, and $\Omega$ is the average angular frequency for an orbit at a certain radius. The inner Lindblad resonance occurs for $l=-1$ and $m=2$, implying $\Omega_\text{p}= \Omega -\kappa/2$. In this case, the star executes two radial oscillations for every rotation of the bar, meaning that it is at the same phase of its oscillation each time an end of the bar swings underneath.

To calculate $\Omega$, we use
\begin{equation}
\Omega = \varv_\text{circ} / r,
\label{eq:omega}
\end{equation}
where $\varv_\text{circ}$ is the circular velocity at a particular radius, $r$. Further to this, following standard theory, we make the approximation that the system is spherically symmetric, and therefore
\begin{equation}
\varv_\text{circ} = \sqrt{G M(\leq r)/ r},
\end{equation}
where $G$ is the gravitational constant, and $M(\leq r)$ is the cumulative mass within radius, $r$. The validity of this approximation is weaker for non-spherically symmetric systems, but previous work has shown that the results have errors of only 5\% - 10\% in the case of more disc-like systems \citep{fragkoudi2021}. This approximation is therefore sufficient for our ends, particularly before the disc-rebuilding process has fully got underway.

Following \citet{kormendy2004}, we calculate $\kappa$ as: 
\begin{equation}
\kappa^2 = 2 \Omega \left(\Omega + \dfrac{\mathrm{d}\varv_\text{circ}}{\mathrm{d}r} \right).
\label{eq:kappa}
\end{equation}

This allows us to calculate the relation $\Omega - \kappa/2$ as a function of radius, where the intersection of this profile with the bar pattern speed provides the location of the inner Lindblad resonance.

The significance of the inner Lindblad resonance lies in its influence on families of stellar orbits. There are two families, in particular, which are important for the formation of bars. In the notation of \citet{contopoulos1980}, these are the $x_1$ orbits, which are elongated parallel to the major-axis of the bar, and the $x_2$ orbits, which have lower eccentricity and are elongated orthogonally to the bar. Stable $x_1$ orbits, naturally, support the formation of a bar, whilst $x_2$ orbits act against it. The domain of each orbit swaps when passing resonant boundaries, with $x_2$ orbits able to exist between the two possible solutions for the inner Lindblad resonance \citep{combes2002}. The result of this is that the larger the range between these solutions is, the more difficult it is for a strong bar to form. This is especially so when the bar is at a nascent stage; when self-gravity is not enough to force orbits to precess at the same rate \citep{kormendy2004}.

With this in mind, the importance of the variations we identified in the mass distribution for the second row of Fig.~\ref{fig:gas_concentration} becomes clear. When we inspect the third row, it can be seen that, at first, the profiles for the inner Lindblad resonance are almost identical, as the cumulative mass profiles at small radii are also similar. However, as the mass concentration in the MHD simulation increases relative to its hydrodynamic analogue, a large divergence takes place. The result of this is that, already by +0.6~Gyr post-merger, an inner Lindblad resonance can exist in MHD simulations for pattern speeds twice as high as in the corresponding hydrodynamic simulation. This situation becomes worse as time proceeds, with the pattern speed required to avoid encountering a broad range of $x_2$ orbits quickly becoming unrealistically high\footnote{These orbits are populated, as can be seen by observing the isotropic star formation during this time evidenced in Fig.~\ref{fig:bar_star_ages} and the edge-on mock images provided in Appendix~\ref{appendix:edge-on-gri-images}.}. Furthermore, as the starburst finishes within the initial 2~Gyr post-merger \citepalias[cf. figure 1 of ][]{whittingham2021}, the inner concentration also undergoes little change after this time. The inner Lindblad resonance therefore stays strong, keeping bar formation in the MHD simulation consistently suppressed post-merger. In hydrodynamic simulations, the pattern speed required to avoid an inner Lindblad resonance is, on the other hand, much more achievable. 

\subsubsection{The formation of a stellar ring in hydrodynamic simulations}
\label{subsec:stellar_ring}

\begin{figure*}
    \includegraphics[width=\textwidth]{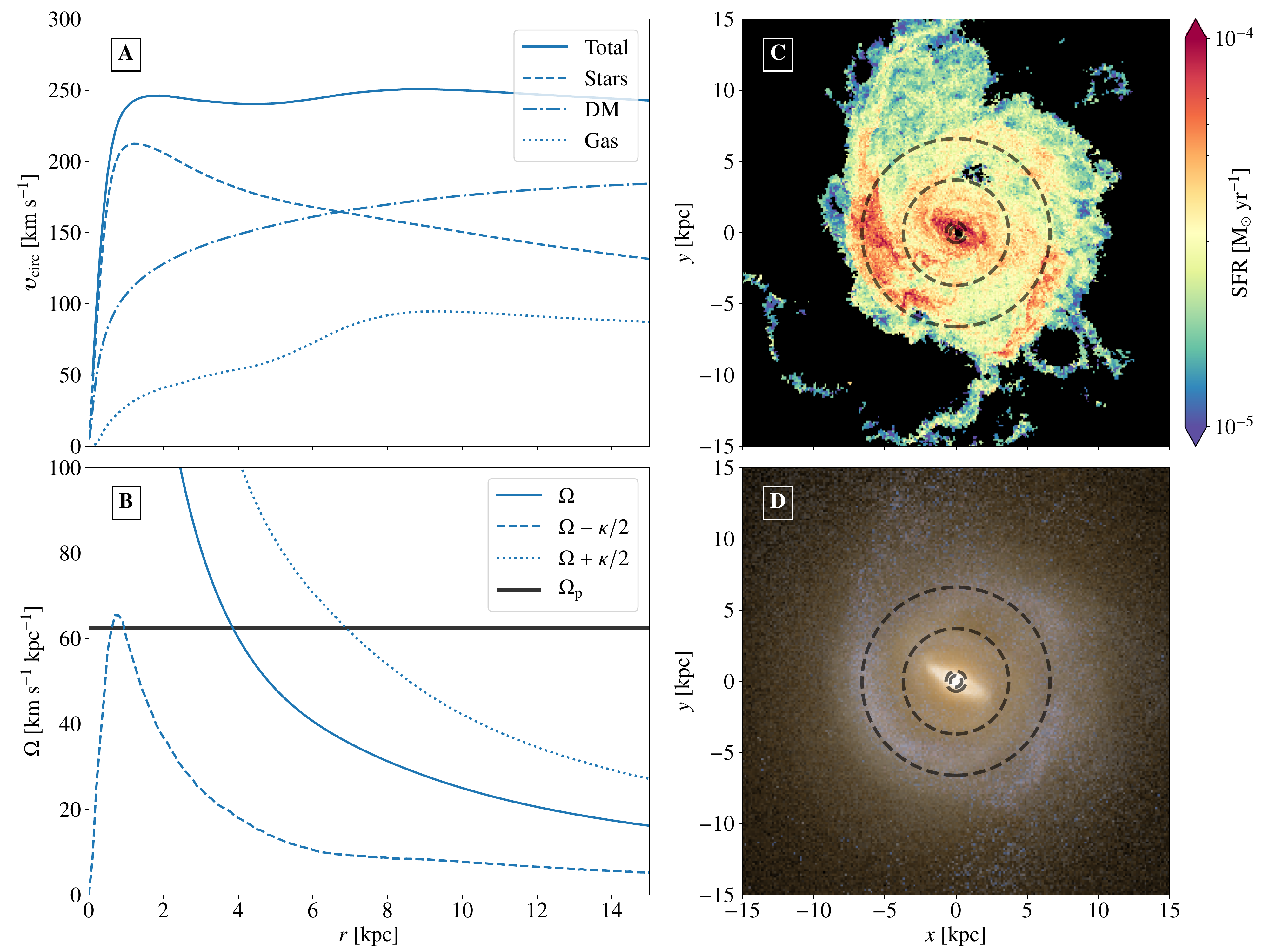}
    \caption{\textit{Top left:} The circular velocity as a function of radius in the disc for 1349-3H at 3 Gyr post-merger. \textit{Bottom left:} The corresponding inner Lindblad (dashed), co-rotation (solid), and outer Lindblad resonances (dotted) as a function of radius. The horizontal line indicates the bar pattern speed measured for this galaxy. The intersection of this line with the profiles marks the radial position of the resonances. \textit{Top right:} A slice through the midplane of the galaxy indicating the star formation rate in each cell, with the resonant positions overlain as dashed circles. \textit{Bottom right:} As above, but the background image now shows a mock \textit{gri} image. Resonances generated by the bar drive gas to the Lindblad resonances, producing a high star formation rate there. This has a pivotal role in how the hydrodynamic remnants develop.}
    \label{fig:resonances}
\end{figure*}

The subsequent growth of a bar in the hydrodynamic simulation has a major impact on how gas and stellar orbits evolve in the remnant. We show evidence of this for 1349-3H in Fig.~\ref{fig:resonances}. We choose to perform this analysis at approximately 3 Gyr post-merger. At this time, the bar has been well-developed for at least a Gyr, and has had a corresponding amount of time to shape the orbits in the disc. Naturally, the pattern speed of the bar varies slightly as it evolves and couples with other modes. At the time we pick, however, the bar pattern speed has varied by no more than $\pm$1 km s$^{-1}$ kpc$^{-1}$ over the last 0.5 Gyr, meaning that the radial positions of the resonances have also stayed approximately constant over the same period. This helps us to better isolate their impact.

In panel~A of Fig.~\ref{fig:resonances}, we show the circular velocity profiles that exist at 3 Gyr post-merger, calculated under the same spherical symmetry assumptions made earlier. The solid line indicates the overall velocity profile, taking into account all matter components. The other lines, meanwhile, take into account only the contribution of stellar, dark matter, and gas components, respectively. It can be seen that stars dominate the dynamics of the central 5~kpc. This is, of course, typical of observed galaxies \citep[see, e.g.,][]{marasco2020}, but it illustrates well why subtle changes in the stellar mass concentration are able to affect the position of the inner Lindblad resonance so strongly. As the cumulative stellar mass increases away from the centre, there is a corresponding rapid increase in the total circular velocity. This increase eventually levels off, with the galaxy maintaining an overall flat rotation profile from then on, as is characteristic of disc galaxies embedded in dark matter haloes. From this point onwards, the $\mathrm{d}\varv_\text{circ}/\mathrm{d}r$ term effectively becomes negligible. Inspecting Eq.~\eqref{eq:kappa}, we see that this leads to $\kappa \propto 1/r$, and therefore also the profile of the inner Lindblad resonance tends to $\Omega - \kappa/2=(1-1/\sqrt{2})\Omega\propto 1/r$. Consequently, if we require the peak of this profile to stay low, the overall velocity curve must begin to flatten later. This, in turn, is only possible if the stellar concentration is kept sufficiently low.

Using Eqs.~\eqref{eq:omega} to~\eqref{eq:kappa}, we obtain the resonant profiles observed in panel~B. In addition to the inner Lindblad resonance, we show two further resonances here: the co-rotation resonance and the outer Lindblad resonance. The condition for the former is fulfilled when an orbit's angular frequency is equal to the forcing frequency: $\Omega = \Omega_\text{p}$. For the outer Lindblad resonance, the condition is $\Omega_\text{p} = \Omega + \kappa/2$ (or $l=1$ and $m=2$, in the language introduced previously) so that the star particle once again performs two radial oscillations for each revolution of the bar, but this time lags behind in the co-rotating reference frame. The presence of these resonances is especially important for gas cells that are not on exactly circular orbits. In this case, gas between the co-rotation and outer Lindblad resonance experiences a net \textit{positive} torque from the bar, whilst that between the co-rotation and inner Lindblad resonance experiences a net \textit{negative} torque \citep{buta1996}. As the resonances are approached, the eccentricity of stellar orbits increases whilst the major axes of the dominating family of orbits rotates by 90 degrees, meaning that orbit crossing becomes inevitable \citep{sellwood1993}. However, the gas cannot interpenetrate and will therefore develop shocks at the position of these resonances, provided its sound speed is not large enough \citep{englmaier1997}, as is the case for the cold star-forming gas. The end effect is that gas is removed from the co-rotation resonance and accumulates at the two Lindblad resonances.

The position of the resonances in the disc can be determined by observing where the resonant profiles intersect with the bar pattern speed, $\Omega_\text{p}$, as indicated by the horizontal, black line in panel~B. This pattern speed was calculated using the standard method based on Fourier decomposition, as applied, for example, in \citet{fragkoudi2021}. We summarise this method in Appendix~\ref{appendix:pattern-speed}. It can be seen that solutions for the inner Lindblad resonance exist at 0.4 and 0.7 kpc, respectively, for the co-rotation resonance at 3.7 kpc, and for the outer Lindblad resonance at 6.6 kpc. As already mentioned, the pattern speed of the bar varies slightly over time. Correspondingly, the radii at which the resonances exist varies over time as well. This is particularly important for the inner Lindblad resonance, which has no solutions when the pattern speed is only a few km s$^{-1}$ kpc$^{-1}$ higher. Overall, the previously-discussed $x_2$ family of orbits is typically restrained to a radial annulus of 0.3 kpc. This is unlikely to be a significant problem for the bar, as we will see in the next two panels.

The impact of the resonances can be understood by examining panel C of Fig.~\ref{fig:resonances}. Here we show a slice through the disc midplane, with colours indicating the star formation rate in each gas cell. Regions in black show where no star formation is taking place. Overlain as dashed circles are the radial positions of the resonances. It can be seen that the regions of heightened star formation align well with the bar near the inner Lindblad resonance and at the edge of the disc near the outer Lindblad resonance. This pattern is a direct tracer of the dense gas that has accumulated at these positions under the action of continuous gravitational torques from the bar. At the outer Lindblad resonance, star formation rates are further increased by the steady accretion of gas post-merger. Indeed, it is possible to see, both above and below the disc, regions of star formation outside the disc. These are dense gas streams, which are helping to fuel star formation in the ring.

\begin{figure*}
    \includegraphics[width=\textwidth]{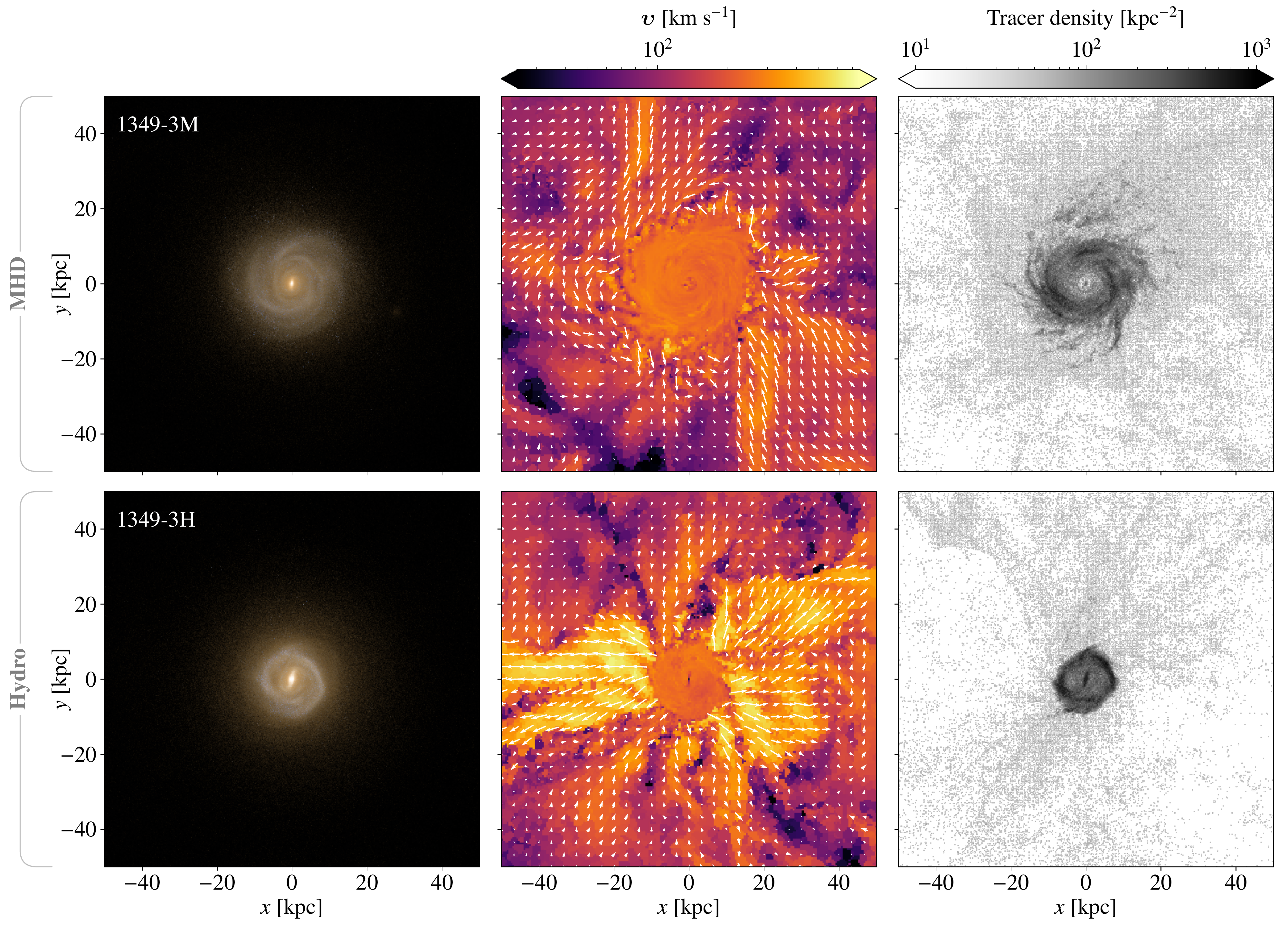}
    \caption{\textit{Left:} Face-on mock \textit{gri} images of the 1349-3 remnants, as seen approximately 5~Gyr post-merger (lookback time of $\sim1.4$~Gyr). \textit{Centre:} The gas velocity in the disc midplane at this time. Arrows indicate the direction, whilst colours indicate the magnitude. We have removed arrows from the approximate area of the disc. \textit{Right:} The surface density of Monte-Carlo tracers that will end up in the disc at $z=0$, where we define this as a cylinder of height $\pm1$~kpc and radius 19.35~kpc (13.14~kpc) for the MHD (hydrodynamic) simulation. In the hydrodynamic simulation, a strong stellar wind disrupts the angular momentum of gas joining the disc, keeping the disc compact. In the MHD run, however, the stellar wind is much less effective, and gas is able to join the disc almost in-situ, helping it to grow rapidly.}
    \label{fig:accretion}
\end{figure*}

The formation of stars in this manner is extremely influential for how the remnant morphology develops. In panel D of Fig.~\ref{fig:resonances}, we show a face-on mock \textit{gri} image of the remnant, created in the same manner as in Fig.~\ref{fig:mock_visual}. Once again, the radial positions of the resonances are overlain. It can be seen the star-forming ring, as indicated by the bluish hues, lines up perfectly with the outer Lindblad resonance. Meanwhile, the co-rotation resonance is practically devoid of new stars, as dense gas has been removed from this region. The bar is also sufficiently large such that the inner Lindblad resonance lies within it. The $x_2$ orbits that would have acted against a weaker bar have therefore almost certainly been subdued by the bar's self-gravity \citep[see, e.g.,][]{kormendy2004}.

Although not presented here, we have performed similar analysis for the simulation Au2-H \citepalias[see figure~9 in][]{whittingham2021} -- a hydrodynamic analogue of one of the original Auriga galaxies. We observe that the star forming ring in this case also aligns with the outer Lindblad resonance. This galaxy is able to grow substantially larger than our merger remnants, however, with a degree of star formation even taking place beyond the outer Lindblad resonance. This is a result of the way accretion takes place in these simulations, and, in particular, the lower star formation rates that result in a more limited impact from wind particles.

\subsection{The impact of stellar feedback on accreting material}
\label{subsec:feedback}

The accretion of gas post-merger, particularly from the former CGM, plays a major role in the rebuilding of a galaxy's disc \citep{sparre2022}. However, as identified in Sec.~\ref{subsec:how_evolution_differs}, the remnants in the hydrodynamic and MHD simulations grow to markedly different sizes. We will show in this section that this predominantly results from the impact of stellar winds on post-merger accretion.

As explained in Sec.~\ref{sec:ISM}, in our simulations, stellar feedback is implemented through the use of wind particles. These are generated at star formation sites and are launched isotropically, interacting only gravitationally until they: a) reach a gas cell with a density that is 5\% of the star formation threshold density, or b) exceed the maximum travel time. At this point the particle's momentum and energy is deposited in its parent gas cell, with energy being split equally into thermal and kinetic parts. In the Auriga simulations, this leads to bipolar winds at late times \citep{grand2017}. This is emergent behaviour arising from the fact that particles encounter lower density gas more quickly when they travel away from the disc midplane; the wind thereafter takes the path of least resistance. In our own simulations, the merger-driven starburst significantly increases the overall number of wind particles formed, helping increase their influence. The result of this is, however, extremely different for the two physics models. We show this in Fig.~\ref{fig:accretion}, where we examine the impact of winds on accretion for the ``1349'' remnants. We do this specifically for a snapshot taken at approximately 5~Gyr post-merger, but our analysis may, of course, be generalised across all simulations and a broad range of times. We show this explicitly in Appendix ~\ref{appendix:stellar_feedback}. 

In the first column of Fig.~\ref{fig:accretion}, we show face-on mock \textit{gri} images, created in the same manner as in Figs.~\ref{fig:mock_visual} and~\ref{fig:resonances}. It can be seen here, that the remnant in the MHD simulation is beginning to form a disc with spiral arms, whilst that in the hydrodynamic simulation has formed the bar and ring morphology previously discussed. These different morphologies lead to a different distribution of wind particles, which alters their impact.

In the second column of the figure, we show a slice through the disc midplane, with colours indicating the magnitude of the gas velocity. Arrows indicate the plane-projected direction of this velocity, with a length scaled to the magnitude of this projection. We have removed arrows from the approximate area of the disc to highlight the dynamics of the CGM. It can be seen that the velocity distributions in each panel exhibit very different patterns; whilst the gas flows in the MHD simulation are predominantly azimuthal, in the hydrodynamic analogue, flows are preferentially radially orientated. These radial outflows are powered by wind particles resulting from the high density star formation at the disc edge, as observed previously in Fig.~\ref{fig:resonances}. As the gas density drops abruptly at the disc edge, wind particles moving in this direction may recouple almost immediately, generating strong, coherent winds, which whisk neighbouring gas away. This leads to a further drop in the gas density at this radius, as was shown quantitatively in \citetalias{whittingham2021}, helping the process to continue.

The strong outflows in the hydrodynamic simulation strongly affect the accretion of gas; because of these, inflows are restricted to areas where star formation -- and therefore the stellar wind -- is weaker, limiting the accretion rate. Moreover, the inflows that do manage to reach the disc are strongly radially-orientated, owing to the disruption of gas angular momentum in the CGM. Together, these factors limit the overall intake of high angular momentum gas in the galaxy, curtailing the growth of the disc. In contrast, star formation in the MHD simulation is spread over a much wider area, with relatively limited star formation at the disc edge. Wind particles are therefore much less effective at disrupting the gas velocity distribution in the CGM and gas that joins the disc can retain its high angular momenta.

\begin{figure}
    \includegraphics[width=\columnwidth]{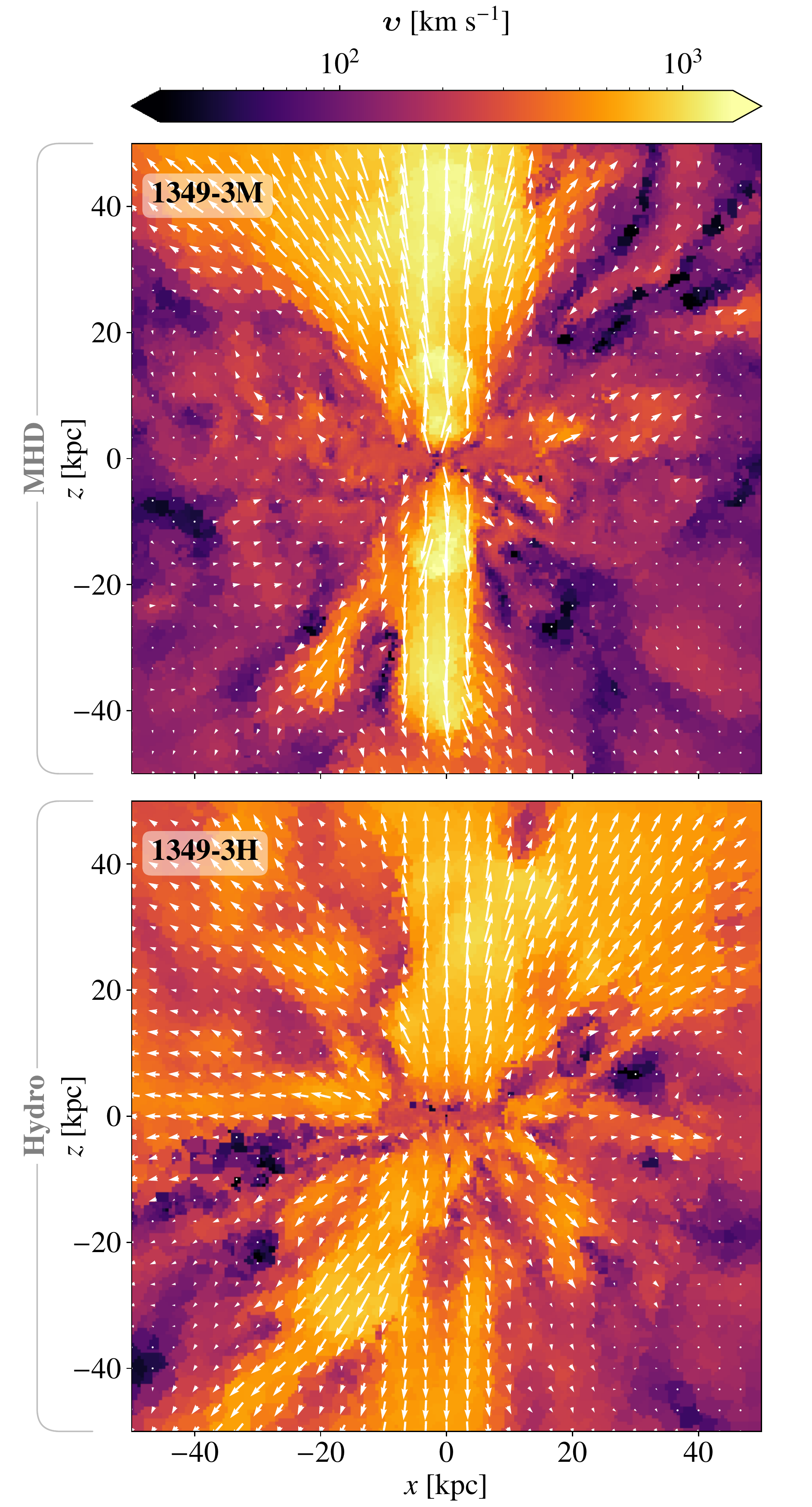}
    \caption{As the second column of Fig.~\ref{fig:accretion}, but showing the remnants edge-on. The strong stellar wind in the hydrodynamic simulation disrupts the CGM in all directions. Meanwhile, in the MHD simulation, outflows are predominantly bipolar and gas in the midplane is consequently able to keep its high angular momentum.}
    \label{fig:outflow}
\end{figure}

\begin{figure*}
    \includegraphics[width=\textwidth]{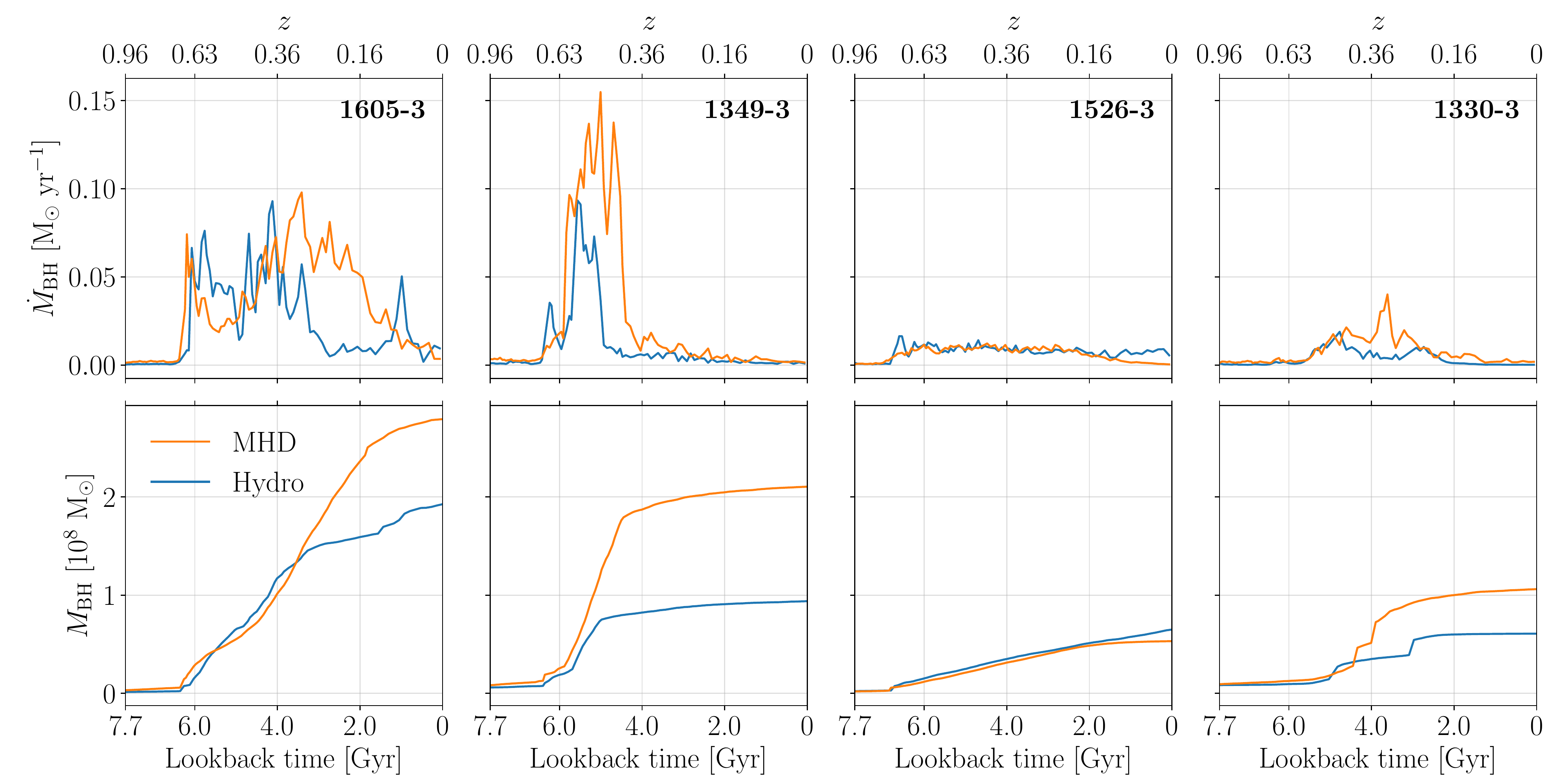}
    \caption{\textit{Top row:} The black hole accretion rate in each simulation as a function of time. \textit{Bottom row:} The black hole mass in each simulation as a function of time. Black holes in MHD simulations can grow up to a factor of 2 larger than their hydrodynamic analogue, owing to the increased gas concentration in these simulations.}
    \label{fig:BH_accretion}
\end{figure*}

We illustrate the differences between how gas accretes onto each remnant in the final column of Fig.~\ref{fig:accretion}. Here we show the surface density of Monte-Carlo tracers (see Sec.~\ref{subsec:arepo_mc_tracers}) that will end up in the disc at $z=0$. We define the disc of each remnant to be a cylinder of depth $\pm1$~kpc with a radius of 19.35~kpc and 13.14~kpc for 1349-3M and 1349-3H, respectively. These radii are the point at which the $B$-band surface brightness drops below $\mu_{B} = 25$ mag arcsec$^{-2}$ \citep[see definition of \textit{optical radius} in][]{grand2017, whittingham2021}. It can be seen that, for the hydrodynamic simulation, a large number of tracers already exist in the bar and ring regions, reflecting the high star formation density here. Outside the disc, however, the density of tracers drops strongly, with tracers only evident in thin filaments, indicating radial accretion of the like identified in the previous column. In the MHD simulation, on the other hand, there is an extensive population of tracers that exist in the immediate neighbourhood of the disc. This population provides a pool of high angular-momentum gas. This joins the disc practically in-situ, thereby enabling its rapid growth.

The full scale of the impact of wind particles can be better understood by also examining the gas dynamics above and below the disc. We show this in Fig.~\ref{fig:outflow} for the 1349-3 remnants for the same time and in the same manner as in the central column of Fig.~\ref{fig:accretion}. It can be seen that, for the hydrodynamic simulation, wind particles dominate the dynamics of practically the entire panel. This is enormously disruptive to the angular momentum of the gas. The distance at which the stellar wind is still active implies that a large-scale fountain flow is in effect, which helps to maintain the radial inflows.

In contrast to this, the velocity distribution in the MHD simulation is predominantly bipolar. This means that gas in the midplane is left mostly unaffected, as is indicated by the arrows, which show extremely small projected velocities. The outflow velocities are generally higher in the MHD simulation by a factor of a few and also originate predominantly from the centre of the disc. This is because, in these simulations, outflows are more greatly influenced by black hole feedback. We explore this in our final analysis section.

\subsection{Altered black hole feedback}
\label{subsec:AGN}

\begin{figure*}
    \includegraphics[width=\textwidth]{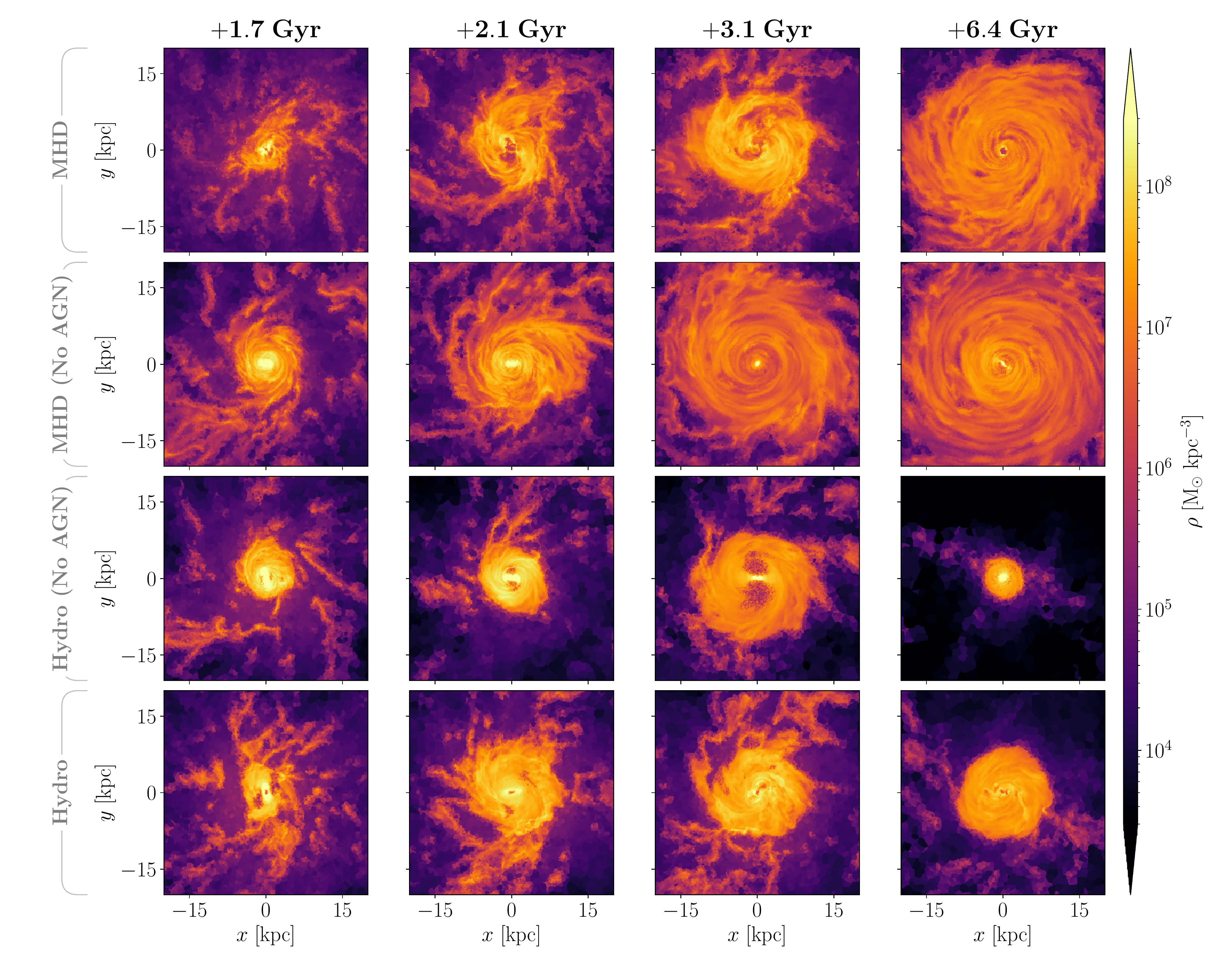}
    \caption{Face-on slices through the disc midplane showing the gas density in the 1349 simulations. Times are given from the start of the merger. \textit{1st row:} Standard MHD simulation. \textit{2nd row:} MHD simulation, but quasar feedback was turned off at the start of the merger. \textit{3rd row:} Hydrodynamic simulation, but quasar feedback was turned off at the start of the merger. \textit{4th row:} Standard hydrodynamic simulation. It is apparent that the morphological differences between hydrodynamic and MHD simulations only become stronger once quasar feedback is removed. Increased quasar feedback in MHD simulations can therefore not be the primary cause of the divergent evolution in the original runs.}
    \label{fig:gas_dist_evolution}
\end{figure*}

In our simulations, as described in Sec.~\ref{sec:ISM}, the energy released by a black hole is directly proportional to its accretion rate. This, in turn, depends on the gas density in the neighbourhood of the black hole \citep[see eq. 8 of][]{grand2017}. As gas is typically more concentrated in our MHD simulations post-merger (see Sec.~\ref{subsec:angmom}), we should expect accretion rates to also be higher and consequently black hole feedback to be more influential. We show that the first of these statements is true in Fig.~\ref{fig:BH_accretion}.

In the first row of the figure, we show the black hole accretion rates for each simulation as a function of time, with MHD simulations shown in orange and hydrodynamic ones in blue. In each case, the arrival of the merging galaxy is associated with an uptick in the black hole accretion rate. Except for 1526-3, it is evident that accretion rates are indeed, on the whole, higher in MHD simulations. The cumulative effect of this increased accretion is that the black hole mass grows substantially larger, as we show in the second row of the figure. In this row, we show the evolution of the total black hole mass as a function of time. Whilst this evolution is dominated by accreted mass, it also includes the impact of black hole mergers. Such mergers produces the discontinuous increases seen, for example, in the 1330-3 simulations. The timing of these increases is different for different physics models, owing to the individual merger trajectories taken, as discussed in Sec.~\ref{subsec:angmom}.

Except in 1526-3, the black hole in the MHD simulation accumulates between 1.5 -- 2 times as much gas as its hydrodynamic analogue by $z = 0$, owing to the increase in baryonic concentration in these simulations. Such density increases will clearly be at their highest in major mergers of gas-rich galaxies. However, under our model, even simulations of more isolated galaxies should exhibit mild density increases when performed with MHD (see Sec.~\ref{subsec:angmom}). These galaxies will therefore also show heightened black hole accretion rates. This implies that increased black hole masses are a generic feature of including magnetic fields in the Auriga model. Nonetheless, even if the average black hole mass increased by a factor of two (i.e. the maximum value seen in Fig.~\ref{fig:BH_accretion}) such values would still be well within the scatter of the well-known black hole  -- halo mass relation \citep{reines2015}. The increase is also clearly only true in a statistical sense; not every remnant in Fig.~\ref{fig:BH_accretion} shows an increase.

The answer as to why the black hole in 1526-3M does not grow larger than its hydrodynamic analogue has already been identified in Sec.~\ref{subsec:angmom}; namely, the magnetic field configuration, and therefore gas density evolution, in this galaxy is different. Here, the magnetic field becomes azimuthally-dominant just as it becomes dynamically important, unlike the non-azimuthal dominance seen in the other three MHD simulations. This means that gas is actually supported from collapse in this simulation, as seen in the angular momentum evolution provided in Fig.~\ref{fig:ang_mom}. Such support may also explain the cessation of black hole accretion in the last $\sim2$~Gyr in this simulation.

Under our black hole model, galaxies that have higher accretion rates necessarily have increased levels of quasar feedback. After the remnant has formed a disc, quasar feedback typically acts to displace gas periodically from the centre. The effect of this can be seen in Fig.~\ref{fig:accretion} through the low tracer density at the centre of the MHD remnant, and in Fig.~\ref{fig:gas_vel-sfr} through the face-on signatures of central outflows and the coincident star formation voids. However, whilst such phenomena are more frequent in the MHD simulations, their impact on the remnant evolution as a whole turns out to be limited. This, perhaps, should be expected, as whilst the black hole accretion rates in Fig.~\ref{fig:BH_accretion} are substantially higher in three out of the four pairs of simulations, morphological differences are observed between \textit{all} pairs of simulations in \citetalias{whittingham2021}; our model must also explain why the 1526-3 simulations evolve differently.

We show explicitly that quasar feedback does not explain the morphological differences in our simulations in Fig.~\ref{fig:gas_dist_evolution}. In this figure, we present a series of slices through the midplane of the 1349-3 simulations showing the gas density. In addition to the standard MHD and hydrodynamic simulations, we also include two further simulations in this figure. In these, we have switched off quasar feedback at the start of the merger (see Fig.~\ref{fig:ang_mom} for times). By doing so, we allow the galaxies to evolve normally pre-merger, and thereby isolate the impact of quasar feedback on the re-growth phase of the disc. The resulting simulation data is naturally not reflective of real galaxies, as, in particular, we remove the pressure support of quasar feedback post-merger, allowing gas to concentrate unphysically at the centre. Nonetheless, the results are instructive. We chose the 1349-3 simulations for this figure as these showed the greatest difference in accretion rates in Fig.~\ref{fig:BH_accretion}, and therefore have the greatest difference in energy output by the black hole post-merger; if quasar feedback is ineffective here, we should not expect it to be effective when the energy output is weaker.

We show the four variations at different times in the process of rebuilding their disc. The physics included in each simulation is labelled on the left-hand side. The amount of time elapsed since the beginning of the merger is also given above each column, with the final column equivalent to $z=0$. In the first column of the figure, the gas discs are all of a similar size. Those simulations where quasar feedback was included appear more disrupted as their morphology has been affected by outbursts, preventing the gas from collapsing neatly into a disc. Such outbursts are particularly strong shortly after the merger, when gas reaches high densities and black hole accretion rates are correspondingly high. 

There are signatures of such outbursts in the top row of Fig.~\ref{fig:gas_dist_evolution} until past the 3 Gyr mark, as evidenced by the density irregularities in the disc until this time. Whilst the most major outbursts take place early on in the rebuilding process, they have a lasting impact. This can be seen by comparing the final disc sizes produced in the two MHD simulations; the disc in the original MHD simulation actually ends up smaller than that in the \textit{MHD (No AGN)} simulation, as outbursts post-merger disrupt the angular momentum of both accreting gas and gas already in the disc. The opposite, however, is true of the hydrodynamic simulations. Here, the final disc size in the \textit{Hydro (No AGN)} simulation is significantly smaller than in the original run. This is because in the hydrodynamic simulations, the dynamics are being more strongly affected by another component; the formation of a central bar.

Both hydrodynamic simulations form bars quickly, but this becomes particularly disruptive in the \textit{Hydro (No AGN)} variation. Here, the bar dominates the centre of the disc, sweeping up gas during its rotation, leading to strong underdensities. Such underdensities are already evident in the +2.1 Gyr snapshot, but are particularly extreme in the following snapshot, where they extend to a distance of a few kpc from the centre. The accretion of gas onto the centre of the bar, however, eventually destroys it, as can be seen in the last snapshot. At this point, support for resonant orbits is removed, and, without any black hole feedback to provide remaining pressure support, the underdensities rapidly fill in, leading to a drop in the disc size.

To summarise, even without quasar feedback, the remnants continue to evolve in ways that are distinctive to the underlying physics models. Indeed, ultimately, the removal of quasar feedback post-merger actually leads to an even larger morphological difference between the two physics models. This suggests that, rather than cause the effect, black hole feedback may actually suppress some of the morphological differences that result from including MHD physics.

\section{Discussion}
\label{sec:discussion}

We have identified four important questions that arise from this work:
\begin{enumerate}
    \item to what extent does the discussed mechanism apply to other mergers?
    \item to what extent does it apply to other galaxy formation models? 
    \item to what extent is the numerical technique used for solving the MHD equations responsible for the results obtained?
    \item how essential are magnetic fields in our model; i.e. does the model rely intrinsically on magnetic fields, or can it be replicated through the tuning of other feedback model parameters?
\end{enumerate}
We attempt to answer these questions below.

\subsection{Applicability of the model to other merger scenarios}

The mergers analysed in this paper are all gas-rich major mergers between disc galaxies situated in MW-sized haloes. Of these properties, it is the gas-rich nature of the mergers that is the most important for our mechanism; firstly, as noted in \citetalias{whittingham2021}, sufficient gas is required in order to amplify the magnetic field through turbulence and adiabatic compression to dynamically-important levels. However, in turn, the magnetic fields in our simulations are also only able to act upon gas, and, as described in detail in Sec.~\ref{subsec:angmom}, it is the motion of this gas in response to torques applied by the magnetic field that ultimately causes the observed morphological differences. We therefore expect magnetic fields to be less influential in gas-poor mergers, such as in the case of mergers between elliptical galaxies.

With this said, virtually all galaxies in cosmological simulations will have undergone a gas-rich merger at some point in their history. Indeed, in \citetalias{whittingham2021}, it was shown that even in the case of isolated, but still cosmological simulations, the consequences of such a merger can be felt for several Gyr after the event. Although a full application of our analysis to such simulations is outside the scope of this paper, we note that our proposed mechanism explains observed features here too, including the appearance of stellar rings at the outer Lindblad resonance (see Sec.~\ref{subsec:stellar_ring}). It seems therefore likely that our mechanism applies more generally in a cosmological context.

\subsection{Applicability of the model to other galaxy formation models}

As described in the introduction of this paper, there are several competing galaxy formation models now available that include an implementation of MHD. However, in only a few of these have magnetic fields been able to impact the dynamics. The inability of the magnetic field to become dynamically important may be linked to a number of factors. For example, it will depend on the seed field strength chosen, the diffusivity of the numerical implementation, and the resolution of MHD phenomena such as amplification through the small-scale dynamo and magnetic draping. As described in \citetalias{whittingham2021}, the magnetic field strengths in Auriga compare favourably with observations of real disc galaxies, which bodes well for analysis of their dynamical importance. We note, too, that simulations where magnetic fields were able to become dynamically important, were able to replicate some of our results. For example, in both \citet{martin-alvarez2020} and \citet{katz2021}, which employed both different numerical methods and implementations of MHD from our own, it was identified that, given sufficiently high field strengths, magnetic fields can torque the gas, thereby reducing the size of the disc, albeit at the expense of using artificially large initial magnetic field strengths (see discussion in Sec.~\ref{sec:numerical_technique}). Such torques form a key part of our own model.

We do not expect a different stellar feedback model to substantially affect the parts of our mechanism that relate to turbulence. For example, whilst the explicit resolution of stellar feedback could generate small-scale turbulence more quickly, thereby shortening the time taken for the magnetic dynamo to reach the non-linear amplification stage, we have already shown in \citetalias{whittingham2021} that this would be unlikely to change the saturation point of the magnetic field. We note as well that whilst stellar-driven turbulence in the CGM and on the ISM-CGM border is already captured by wind particles \citep{pakmor2020, vandevoort2021}, during the merger, turbulence in the ISM is overwhelmingly gravitationally-driven \citepalias[see, e.g., fig.\ 15 of][]{whittingham2021}. This is already fully-captured in the \citet{springel2003} ISM model.

With this said, other aspects of stellar feedback could still play a significant role. For example, more explosive stellar feedback would likely disrupt the formation of high density structures, having a particularly strong impact on the formation of stellar rings. In contrast, the wind particle implementation, as used in Auriga, allows gas to stay at high densities, as the multiphase nature of the ISM cannot be resolved and wind particles only recouple below a threshold density. Indeed, it is noticeable that in the original hydrodynamic Illustris simulation, which also used a wind particle implementation, galaxies frequently formed ring like structures \citep[see, e.g., fig.\ 1 and 13 of][]{marinacci2014}. Wind particles in this simulation were launched with a bipolar model, where particles were explicitly launched away from the disc \citep[see, e.g.][for further details]{pillepich2018}. However, as seen in Fig.~\ref{fig:outflow}, this would likely be effective enough to disrupt the angular momentum of the CGM, as required under our model. The updated Illustris TNG model \citep{nelson2019}, meanwhile, forms approximately the right frequency of barred galaxies \citep{zhao2020} and no longer forms such a large number of disc galaxies with star-forming rings (c.f.\ fig.\ 6 of \citealt{snyder2015} and fig.\ 5 of \citealt{rodriguez-gomez2019}). One of the major advances made in Illustris TNG compared to the original Illustris model was the implementation of magnetic fields. The importance of this addition may not have been fully appreciated.

Finally, we expect the cosmological nature of the simulation to play a large role. This affects many aspects of galaxy evolution. For example, as shown in \citet{sparre2022}, a substantial fraction of star formation post-merger originates from gas that was previously outside the discs of the progenitors. Without this additional gas, star formation rates would be lower, thereby reducing the impact of winds and the ability of the magnetic field to affect the disc rebuilding process. The existence of such gas also helps to maintain turbulence in the galaxy, aiding the amplification of the magnetic field, and therefore its dynamical importance, as examined in \citetalias{whittingham2021}. Furthermore, isolated simulations of galaxies are typically initialised with the magnetic field in an almost purely azimuthal or toroidal configuration. In contrast, in our own cosmological simulations, we find that the magnetic field can exhibit strong non-azimuthal components. Indeed, these are vital for producing the increased baryonic concentrations identified in Sec.~\ref{subsec:angmom}. This points to the need to model magnetic fields self-consistently.

\subsection{Requirement on the numerical technique for resolving magnetic field growth}
\label{sec:numerical_technique}

Cosmological magnetic fields are believed to have grown from seed fields produced in the early Universe. Typically, one of two sources are invoked for the production of such seeds: i) the Biermann battery mechanism, which is able to generate magnetic fields in proto-galaxies with typical values of $10^{-20}$ Gauss and coherence scales of several kpc, and ii) primordial magnetic fields, which could be produced with similar strengths during the epoch of cosmic inflation or during phase transitions in the post-inflation era \citep{widrow2002,kulsrud2005,brandenburg2005}. By adopting seed fields of such strength and modelling amplification in a small-scale dynamo with realistic Reynolds numbers of order $\rmn{Re}\sim10^{11}$, it is possible to theoretically explain the micro-Gauss strength of magnetic fields observed in galaxies today \citep{schober2013}. 

It turns out, however, that simulating this process explicitly is extremely computationally challenging. Indeed, current-day galaxy formation simulations are still far from resolving the necessary scales of turbulence required to amplify the magnetic field in the requisite time frame. As a result, the strength of the seed field must be artificially increased in order to make up for the missing resolution. However, at the same time, care must be taken to prevent increasing it to the point that the subsequent magnetic field unphysically modifies the process of galaxy formation e.g., by preventing gas accretion onto the forming disc through dynamically important magnetic pressure resulting from the adiabatically compressed field  \citep{marinacci2016,martin-alvarez2020,katz2021}.

Three possibilities exist to circumvent the aforementioned problems. Firstly, adaptive mesh-refinement simulations of magnetic field growth in galaxies can adopt extremely small (quasi-uniform) resolutions in the high-density regions of interest to be able to produce magnetic fields at the observed strengths \citep{martin-alvarez2022}. This method is, however, currently only appropriate for cosmological simulations of galaxies forming in isolation. Alternatively, the effective resolution can be increased by introducing a turbulent subgrid scheme, where the magnetic field growth via the small-scale dynamo is modelled below the formal grid resolution. This avoids the otherwise large numerical diffusion at the grid scale, which would preclude simulating a magnetic dynamo \citep{liu2022}. This approach, however, necessarily requires the addition of more free parameters to the overall model, which must then be tuned. The final approach is to use a moving mesh code. Because the numerical truncation error of a given numerical scheme is proportional to the sum of the absolute values of sound speed and gas velocity relative to the mesh, the numerical diffusion can be substantially reduced and the effective Reynolds number thus increased by using a Voronoi mesh that is co-moving with the flow \citep{springel2010,bauer2012}. By using this method, we substantially boost the effective resolution, enabling us to resolve the small-scale dynamo in galaxies, whilst ensuring that the magnetic fields do not artificially interfere with the collapse and formation of the galaxy \citep{pakmor2017,pfrommer2022}.

\subsection{Can the effect of our model be mimicked in hydrodynamic simulations?}

Despite the broad range of differences between galaxy formation models, each claims to be able to replicate some aspect of galaxy evolution. This implies a certain level of degeneracy in these models, given the current level of observational error attached. It is therefore natural to ask: are magnetic fields actually required for creating accurate galaxies in Auriga, as proposed here, or can their impact be replicated by another mechanism? The most likely candidate for this would be the feedback implementation, given its well-documented impact on star formation processes. We note, for example that recent work has shown that quasar feedback may help to weaken bars in Auriga \citep{irodotou2022}. This would help to reduce the likelihood of forming a star-forming ring under our model. As explained in Sec.~\ref{subsec:AGN}, however, the overall impact is unlikely to be enough. The impact of more influential black hole feedback in a still hydrodynamic model can, furthermore, be observed in our own simulations in 1526-3H (see Appendix B of \citetalias{whittingham2021}). As can be seen in fig.~7 of \citetalias{whittingham2021}, whilst this does indeed weaken the bar, the remnant still shows a substantially different morphology compared to its MHD analogue.

Stellar feedback has also been shown to be highly influential in merger simulations for a range of models \citep[see, e.g.,][]{moreno2019, moreno2021, li2022}. However, rescaling our stellar feedback would, too, almost certainly not prevent the observed morphological divergence. This can be seen through inspection of the MHD and hydrodynamic versions of the Auriga simulations, as shown in \citetalias{whittingham2021}. For these galaxies, star formation was not as intense, and subsequently fewer wind particles were generated, reducing their effectiveness. On the one hand, this meant that the CGM was less disturbed and so the remnants could grow larger. Ultimately, however, a similar morphological divergence still takes place; hydrodynamic simulations still exhibit bar-and-ring structures whilst the remnants in the MHD simulations are predominantly MW-like.

More fundamentally, feedback and magnetic fields act in different ways; whilst feedback can transport the angular momentum of gas to large galactocentric radii, magnetic fields are able to promote inwards transport via magnetic draping \citep{lyutikov2006,dursi2008,pfrommer2010,berlok2019}, before magnetic tension forces transport and redistribute the angular momentum locally. This is inherently different and allows magnetic fields to initially reduce the size of the disc before helping to grow it substantially. In contrast, feedback through disruption can only reduce the size of the disc. We conclude from this that feedback can neither be tuned nor modified to replicate the mechanism we have presented in this paper.

\section{Conclusions}
\label{sec:conclusions}

In this paper, we have investigated how magnetic fields are able to affect galaxy mergers in the framework of the Auriga galaxy formation model. To do this, we have analysed the simulations first presented in \citetalias{whittingham2021}. These are a series of high-resolution (dark matter resolution equal to $1.64 \times 10^5 \; \mathrm{M}_\odot$) cosmological zoom-in simulations of major mergers between disc galaxies in MW-like haloes. The mergers take place between $z=0.9-0.5$, and all remnants are subsequently able to regrow a disc. The remnant disc, however, is systematically larger in MHD simulations and also shows spiral arm features and an extended radial profile. In contrast, in hydrodynamic simulations, the remnant is compact and displays prominent bar and ring components. We have presented a mechanism in this paper that explains how magnetic fields cause this morphological divergence. Our model is provided as a schematic in Fig.~\ref{fig:schematic} and is as follows:

\begin{enumerate}
    \item Within a few 100~Myr of the first closest approach, the magnetic field becomes dynamically dominant. Non-azimuthally orientated parts of the field then effectively redistribute angular momentum between accreting gas and the gas in the disc. When the field is predominantly non-azimuthally orientated, this leads to an initial reduction in the disc size (Figs.~\ref{fig:mock_visual}, ~\ref{fig:bar_star_ages} and~\ref{fig:ang_mom}).
    \item The resultant higher baryonic concentration produces a strong inner Lindblad resonance, which suppresses the formation of a bar. When the magnetic field is predominantly azimuthally-orientated, the support it provides against collapse performs the same role. In the hydrodynamic runs, however, a large bar forms easily (Figs.~\ref{fig:mock_visual}, \ref{fig:bar_star_ages}, \ref{fig:gas_concentration}, and \ref{fig:resonances}).
    \item In the hydrodynamic simulation, the large bar shepherds gas towards the outer Lindblad resonance, resulting in a high star formation rate in this region. The absence of a strong bar in the MHD simulation, on the other hand, allows the gas to remain flocculent and for spiral arm features to develop (Figs.~\ref{fig:mock_visual},  \ref{fig:bar_star_ages}, and \ref{fig:resonances}). 
    \item The high star formation rate density in the hydrodynamic simulation launches a strong stellar wind away from the disc, disrupting the angular momentum of neighbouring gas cells, thereby keeping the remnant compact. In contrast, in the MHD simulation, winds are less effective and gas on the outskirts of the disc retains much of its angular momentum, resulting in rapid disc growth (Figs.~\ref{fig:accretion} and~\ref{fig:outflow}).
\end{enumerate}

In addition, we also find in this paper that:
\begin{itemize}
    \item Torques provided by the magnetic field are able to systematically reduce the time taken until coalescence (Fig.~\ref{fig:ang_mom}). This effect is particularly strong for inspiralling mergers, which experience several fly-bys.
    \item The increased gas concentration in MHD simulations is able to grow the central black hole up to a factor of two greater than in the hydrodynamic analogue (Fig.~\ref{fig:BH_accretion}). The subsequent increase in quasar feedback. however, does not have a significant impact on the remnant evolution (Fig.~\ref{fig:gas_dist_evolution}).
 \end{itemize}
 
Whilst the impact of magnetic fields is probably strongest under our set-up, we have shown in \citetalias{whittingham2021} that this impact is also felt in more isolated, but still cosmological galaxy simulations. Furthermore, as discussed in Sec.~\ref{sec:discussion}, it seems highly unlikely that this impact could be replicated by a different feedback mechanism. We therefore conclude that magnetic fields are a crucial element of modelling galaxy formation in a cosmological environment, and that the modelling of disc galaxies, in particular, cannot be done correctly in purely hydrodynamic simulations.

\section*{Acknowledgments}

We would like to thank Francesca Fragkoudi, Freeke van de Voort, and Dylan Nelson for stimulating conversations at the Virgo 2022 conference. JW acknowledges support by the German Science Foundation (DFG) under grant 444932369. MS and CP acknowledge support by the European Research Council under ERC-CoG grant CRAGSMAN-646955 and ERC-AdG grant PICOGAL-101019746.

\section*{Data Availability}

The data underlying this article will be shared on reasonable request to the corresponding author.

\bibliographystyle{mnras}
\bibliography{main}

\begin{appendix}

\section[Appendix A: Edge-on mock stellar light images]{Analysis of edge-on mock stellar light images}
\label{appendix:edge-on-gri-images}
\begin{figure*}
    \includegraphics[width=\textwidth]{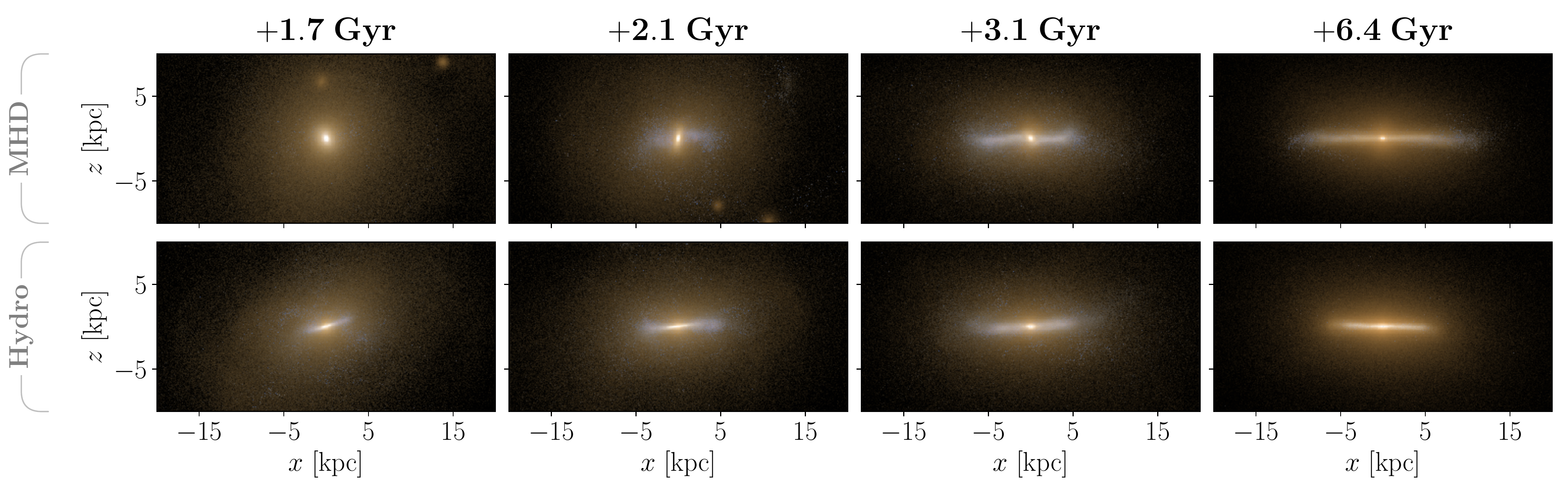}
    \caption{As Fig~\ref{fig:mock_visual}, but the images are now shown edge-on. In the MHD simulation, the growth of the young stellar disc takes place in misalignment with the bar. This is caused by the magnetic field torquing the gas disc.}
    \label{fig:mock_visual-edge-on}
\end{figure*}

In Fig.~\ref{fig:mock_visual-edge-on}, we show edge-on mock \textit{gri} images of the 1349-3 remnants as they evolve post-merger. These images are created in the same way as in Fig.~\ref{fig:mock_visual} and are shown at the same times. The images are orientated with the cold, dense gas disc, which, especially for the hydrodynamic simulation early on, does not always fully line up with the young stellar disc. This is due to the fact that the accreting gas has angular momentum  misaligned with said disc.

Several aspects in Fig.~\ref{fig:mock_visual-edge-on} support the analysis already provided in Sec.~\ref{subsec:how_evolution_differs}. For example, the remnant in the MHD simulation at 1.7~Gyr post-merger is more compact than its hydrodynamic analogue. Similarly, the radial growth of the remnant in the hydrodynamic simulation stalls over time, whilst in the MHD simulation, growth is rapid following the initial compaction stage. The thickness of each disc at +6.4~Gyr ($z=0$) is similar, although in the hydrodynamic simulation the disc ends up marginally thinner. This originates from the thinner gas disc \citepalias[see fig.\ 6 of][]{whittingham2021}, which in turn arises from the stronger stellar winds in this galaxy, as analysed in Sec.~\ref{subsec:feedback}.

Where the remnants differ in particular, however, is in the accretion of material. For example, it is clear in the second column of Fig.~\ref{fig:mock_visual-edge-on} that in the MHD simulation the bar is misaligned with the young stellar disc. This is in contrast to the bar in the hydrodynamic simulation, which sits aligned with this disc. This misalignment is a direct result of the magnetic field applying torques to the reforming gas disc, as we analysed in Sec.~\ref{subsec:angmom}. This reduces the influence of the bar and allows stars to be born very close to the centre, thereby populating the $x_2$ orbits discussed in Sec.~\ref{subsec:resonances}. Such orbits will ultimately suppress the bar, causing it to become noticeably smaller, as can be observed in Fig.~\ref{fig:mock_visual}.

\section[Appendix B: Generalising the stellar feedback analysis to all simulations]{Generalising the stellar feedback analysis to all simulations}
\label{appendix:stellar_feedback}

\begin{figure*}
    \includegraphics[width=\textwidth]{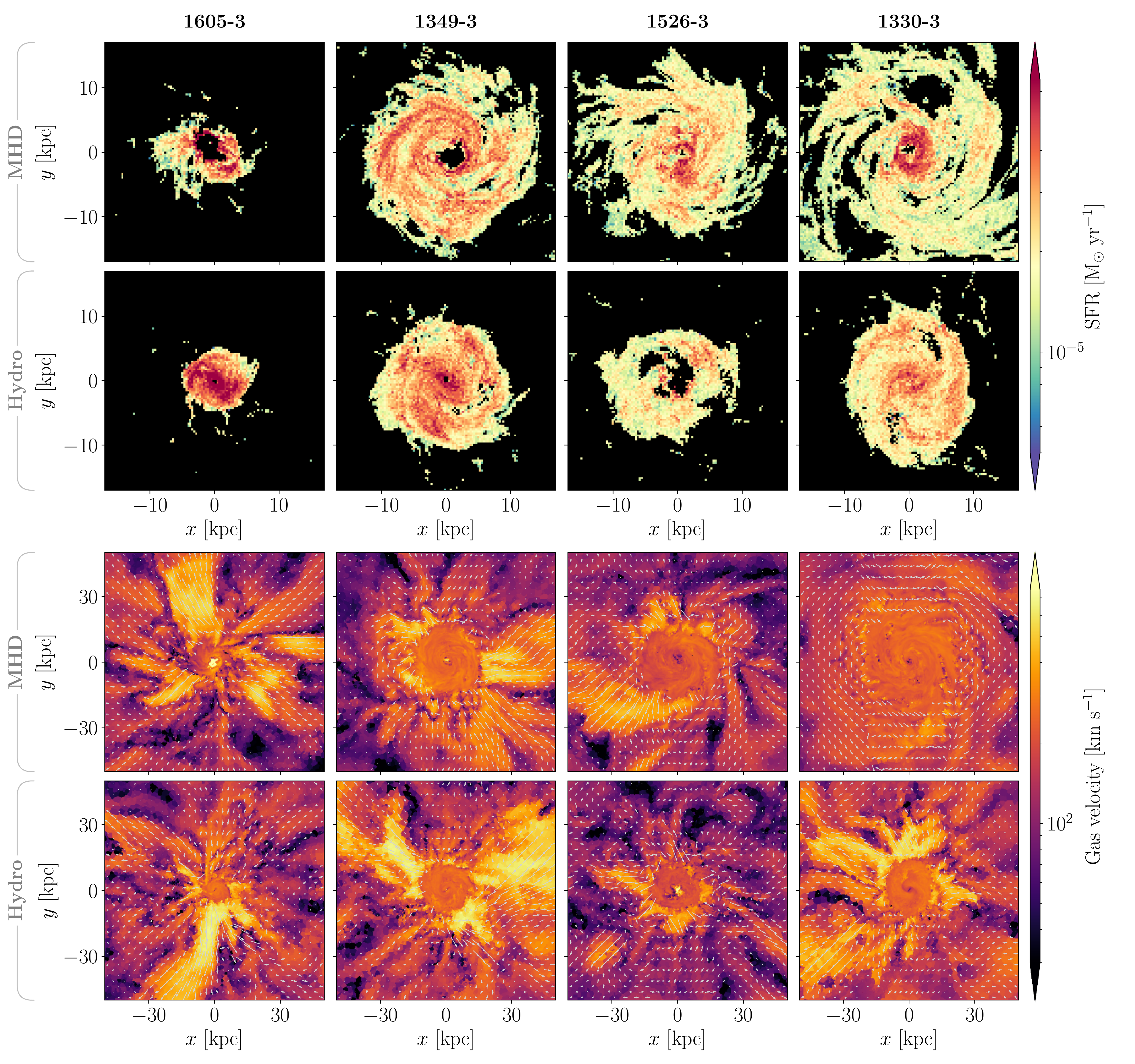}
    \caption{\textit{1st and 2nd row:} Face-on slices through the midplane of the disc for MHD and hydrodynamic simulations, respectively, where colours indicate the star formation rate in each cell. Remnants are seen at +4 Gyr after the beginning of the merger. \textit{3rd and 4th row:} As above, but colours indicate gas velocity, with arrows indicating the projected velocity in the CGM. The bar-and-ring structure observed for 1349-3M in earlier case studies is typical of all hydrodynamic simulations in our suite. The star forming rings strongly disrupt the dynamics of the neighbouring gas, preventing the accretion of high angular momentum gas and keeping the remnants compact. Star formation in MHD simulations, on the other hand, is more evenly distributed and the CGM is subsequently less disrupted, helping the remnant discs to grow faster radially.}
    \label{fig:gas_vel-sfr}
\end{figure*}

During this paper, we have frequently used the 1349-3 simulations as a case study. In this section, we support our claim that conclusions drawn from these studies may be generalised to the wider simulation suite. We focus, in particular, on our assertion that star formation is distributed differently in MHD simulations and that this, in turn, produces a velocity distribution in the CGM that is more conducive to the growth of the disc. We show this with the aid of Fig.~\ref{fig:gas_vel-sfr}. In the top two rows of the figure, we show the star formation rate distribution, as previously displayed in panel C of Fig.~\ref{fig:resonances}. In the bottom two rows, we show the gas velocity distribution for all simulations, displayed in the same manner as in Figs.~\ref{fig:accretion} and~\ref{fig:outflow}. Each panel shows a face-on slice through the galactic midplane, as observed at 4~Gyr post-merger.

In the upper panels, it can be seen that star formation in the MHD simulations takes on a complicated structure, reflecting the underlying flocculent gas distribution. Star formation at the edge of the disc, in particular, is strongly inhomogeneous, owing to the increased effects of stochasticity that arise with decreasing density. Star formation is, in general, distributed throughout the disc. However, some coherent structures are also evident, such as the spiral arms in 1349-3M and 1526-3M, and the bulge elements in 1526-3M and 1330-3M. Star formation in the hydrodynamic simulations, on the other hand, is typically distributed in bar and ring structures, as previously discussed in Sec.~\ref{subsec:resonances}. Signs of this are clearly visible in three of the galaxies (1605-3H, 1349-3H, and 1330-3H). In 1526-3H, the central black hole is unusually active (see appendix B in \citetalias{whittingham2021}), resulting in a star formation void at the centre. Even here, however, there are weak signs of the bar-and-ring structure. Indeed, this structure is also visible in the optical counterpart of the galaxy \citepalias[see figure~7 of][]{whittingham2021}, albeit less clearly defined compared to the other hydrodynamic remnants.

In the lower panels of Fig.~\ref{fig:gas_vel-sfr}, it can be seen that the differences in star formation distribution generally translate to a different CGM velocity distribution. In particular, outflows are typically stronger in the hydrodynamic simulations, resulting in a more disturbed velocity distribution. The differences between the physics models are greatest when the merger scenario was most inspiralling, as is the case in the 1330 simulations. Such mergers lead to a greater retention of high angular momentum gas, helping the disc to grow more quickly. In MHD simulations, this spreads star formation over a larger area, reducing the effectiveness of wind particles. Indeed, inspection of the gas velocity distribution for 1330-3M shows that gas at the edge of the disc is rotating almost perfectly azimuthally. This in strong contrast with its hydrodynamic analogue. For mergers that were more ``head-on'', star formation rates are higher and disc growth is slower, as the CGM consists of lower angular momentum gas. Both of these factors increase the star formation rate density. This results in more effective winds and stronger disruption of the CGM. Even in this scenario, however, winds are, overall, more effective in hydrodynamic simulations, owing to the strong star-forming rings formed. These launch significantly faster outflows, which penetrate further into the CGM. Such outflows are better able to disrupt the CGM, reducing the amount of high angular momentum gas still further, thereby helping to keep the disc compact. 

Winds are stronger in the MHD simulations shown here compared to that shown in Fig.~\ref{fig:accretion} as the disc is still relatively young and star formation is still comparably high at this point in time. Choosing such a time was necessary in order to better highlight the star formation distribution in the upper two rows. It is noticeable, however, that just 1~Gyr later in Fig.~\ref{fig:accretion}, the CGM in 1349-3M has returned to a predominantly azimuthal velocity distribution, whilst the distribution in the hydrodynamic case is still highly disturbed. This shows how enduring the impact of the bar-and-ring structure is on the evolution of the hydrodynamic remnant and its environment.

\section[Appendix C: Calculating the bar pattern speed]{Calculating the bar pattern speed}
\label{appendix:pattern-speed}

To calculate the bar pattern speed, we take Fourier decompositions of stellar surface density projections centred on the bar potential minimum and analyse the evolution of the symmetric $m=2$ mode. In quantifiable terms, this means we calculate the components:

\begin{equation}
\label{eq:a_m}
    a_m(r) = \sum_{i=0}^N M_i \cos({m \theta_i}),
\end{equation}
\begin{equation}
\label{eq:b_m}
    b_m(r) = \sum_{i=0}^N M_i \sin({m \theta_i}),
\end{equation}
where $m$ is the Fourier mode, $r$ is the projected distance taken in cylindrical bins of 0.25 kpc from the centre, $N$ is the number of star particles in each bin, $M_i$ is the mass of an individual star particle, $i$, and $\theta_i$ is its azimuthal angle in the plane of the disc. We include all star particles within $\pm5$ kpc of the midplane for this calculation.

Using these components, we find the radial value at which the normalised bar strength, $A_2(r)$, reaches its peak, where this is calculated as:
\begin{equation}
    A_2(r) = \left( \frac{\sqrt{a_2(r)^2 + b_2(r)^2}}{a_0(r)} \right).
\end{equation}
Evaluating the components $ a_2(r)$ and $ b_2(r)$ at this peak radius, $r_\rmn{max}$, we can then calculate the angle at which the $m=2$ mode, and therefore the bar, lies in the plane of the disc:
\begin{equation}
   \theta = 0.5 \arctan{ \left(\frac{b_2(r_\text{max})}{a_2(r_\text{max})} \right)}.
\end{equation}
The instantaneous bar pattern speed is then simply:
\begin{equation}
   \Omega_\text{p} = \frac{\Delta \theta}{\Delta t},
\end{equation}
where $\Delta t$ is the time between angle calculations. For the time at which we perform our analysis, we have sufficient snapshot cadence such that pattern speeds may be recovered unambiguously. Nonetheless, we have also calculated the bar pattern speed using the Tremaine-Weinberg method \citep{tremaine1984} in order to cross-check our calculated values. This method returns similar values within an approximate error margin of 5 km$^{-1}$kpc$^{-1}$. This is within the expected accuracy bounds of this method \citep[see, e.g.,][]{fragkoudi2021}.

\end{appendix}

\bsp	
\label{lastpage}
\end{document}